\documentclass[epj]{svjour}
\usepackage{latexsym}
\usepackage{graphics}
\usepackage{amsmath}
\usepackage{amssymb}
\usepackage{amsmath,bm}
\usepackage{graphicx}
\usepackage[normalem]{ulem}
\usepackage[dvips]{color}
\renewcommand\sout{\bgroup \color{red} \ULdepth=-.5ex \ULset}

\begin{document}
\authorrunning{Lie-Wen Chen {\it et al.}}
\titlerunning{Probing isospin- and momentum-dependent nuclear effective interactions}
\title{Probing isospin- and momentum-dependent nuclear effective interactions in neutron-rich matter}
\author{Lie-Wen Chen\inst{1,2}, Che Ming Ko\inst{3}, Bao-An Li\inst{4,5}, Chang Xu\inst{6}, and Jun Xu\inst{7} 
}                     
%
%
\institute{ Department of Physics and Astronomy and Shanghai Key
Laboratory for Particle Physics and Cosmology, Shanghai Jiao Tong
University, Shanghai 200240, China \and Center of Theoretical
Nuclear Physics, National Laboratory of Heavy Ion Accelerator,
Lanzhou 730000, China \and Cyclotron Institute and Department of
Physics and Astronomy, Texas A$\&$M University, College Station, TX
77843-3366, USA \and Department of Physics and Astronomy, Texas
A$\&$M University-Commerce, Commerce, TX 75429-3011, USA \and
Department of Applied Physics, Xi'an Jiao Tong University, Xi'an
710049, China \and Department of Physics, Nanjing University,
Nanjing 210008, China \and Shanghai Institute of Applied Physics,
Chinese Academy of Sciences, Shanghai 201800, China }
\date{Received: date / Revised version: date}
%
\abstract{The single-particle potentials for nucleons and hyperons
in neutron-rich matter generally depends on the density and isospin
asymmetry of the medium as well as the momentum and isospin of the
particle. It further depends on the temperature of the matter if the
latter is in thermal equilibrium. We review here the extension of a
Gogny-type isospin- and momentum-dependent interaction in several
aspects made in recent years and their applications in studying
intermediate-energy heavy ion collisions, thermal properties of
asymmetric nuclear matter and properties of neutron stars. The
importance of the isospin- and momentum-dependence of the single-particle
potential, especially the momentum dependence of the isovector potential,
is clearly revealed throughout these studies.
} 
\maketitle

\section{Introduction}\label{introduction}

One of the fundamental questions in contemporary nuclear physics and
astrophysics is to understand quantitatively the in-medium nuclear
effective interactions, which are directly related to the structure
and decay properties of finite nuclei, the reaction dynamics induced
by nuclei, the equation of state (EOS) of dense nuclear matter, the
properties of compact stars, and the explosion mechanism of
supernova~\cite{LiBA98,LiBA01b,Lat00,Dan02a,Bar05,Ste05a,CKLY07,LCK08}.
The in-medium nuclear effective interactions generally depend on the
medium baryon density and isospin asymmetry, the particle momentum,
and the particle isospin. Theoretically, information on the in-medium
nuclear effective interactions usually can be extracted from various
approaches using the microscopic many-body theory, the effective-field
theory, and phenomenological models. In the microscopic many-body theory approach~\cite{Bru67,Sie70,Sjo74,Bom91,Zuo02,Har87,Mut00,Dic04,Fri81,Lag81,Akm98},
vacuum bare nucleon-nucleon (NN) interactions, that are fitted to
high-precision experimental data, are used to describe the nuclear system,
and  the resulting in-medium nuclear effective interactions (G-matrix) are
therefore free of parameters. In the effective-field theory (EFT)
approach~\cite{Ser97,Fur04,Vre04}, effective nuclear interactions are
constructed from low energy QCD and its symmetry breaking, and thus they
usually have a smaller number of free parameters and a correspondingly
higher predictive power. In the phenomenological
approach~\cite{Ser86,Chi77,Rei89,Rin96,Vau72,Bra85,Sto07,Tre86,Ban90,Bas03,Bas05,Muk07,Bas08,Bas09,Cho09,Rou11,Rou12,Gai13,Bas13},
the nuclear many-body problem is generally treated using effective
density- (momentum- and isospin-) dependent nuclear forces or effective
interaction Lagrangians with parameters adjusted to fit the properties of
a number of finite nuclei and/or nuclear scattering data. Although a lot of
useful information on the in-medium nuclear effective interactions around
normal nuclear density with relatively small isospin asymmetry has been
obtained by analyzing experimental data based on these various approaches,
corresponding knowledge at high baryon density and/or large isospin asymmetry
remains very limited.

Indeed, both the EFT and phenomenological approaches usually give
excellent descriptions of the nuclear matter properties around or below
the saturation density with relatively small isospin asymmetry. Their
predictions on properties of nuclear matter at the high density region
and/or large isospin asymmetry are, however, likely unreliable. In
addition, due to different approximations or techniques adopted in
various microscopic many-body theory approaches, their predictions on
the properties of nuclear matter and the in-medium nuclear effective
interactions could be very different even for the same vacuum bare
NN interaction~\cite{Die03,LiZH06}. In particular, predictions on
the isovector properties of nuclear matter, especially the density
dependence of the nuclear symmetry energy, are still largely
different in different microscopic many-body theory approaches or
the same microscopic many-body theory approach but with different
vacuum bare NN interactions~\cite{LiZH06}. Therefore, the experimental
information on the in-medium nuclear effective interactions and
properties of nuclear matter at extremely large isospin and high
baryon density is of critical importance, and this provides a strong
motivation for studying isospin-dependent phenomena with radioactive
nuclei at the new/planning rare isotope beam facilities around the
world, such as CSR/Lanzhou and BRIF-II/Beijing in China, RIBF/RIKEN
in Japan, SPIRAL2/GANIL in France, FAIR/GSI in Germany, SPES/LNL in
Italy, RAON in Korea, and FRIB/NSCL and T-REX/TAMU in USA.

In the present contribution, we review the isospin- and
momentum-dependent MDI interaction, which has been
extensively used in recent years to study heavy ion collisions
induced by neutron-rich nuclei, the thermal properties of asymmetric
nuclear matter including its liquid-gas phase transition,
and the properties of neutron stars. We also review the
extended MDI interaction for the baryon octet and its application
to hybrid stars as well as the improved MDI interaction with separate
density dependence for neutrons and protons which takes into
account more accurately the effect of the isospin dependence of
in-medium NN interactions. We further highlight some results from
these studies with emphasis on the effects of the momentum dependence
of nuclear mean-field potential, especially its isovector symmetry
potential in asymmetric nuclear matter. These studies demonstrate
the importance of both the isospin and momentum dependence of nuclear
mean-field potential in asymmetric nuclear matter, and provide
useful information on the isospin- and momentum-dependent effective
interactions in such a matter.

This article is organized as follows. In Section~\ref{SectionIsospin},
we briefly introduce the nuclear symmetry energy and the nuclear symmetry
potential. Section~\ref{SectionMDI} gives some details of the isospin-
and momentum-dependent MDI interaction. We then highlight some
applications of the MDI interaction in Section~\ref{SectionApplications}.
In Section~\ref{SectionExtension}, the extended MDI interaction for the
baryon octet and the improved MDI interaction with separate density
dependence for neutrons and protons are reviewed. Finally, a summary is
presented in Sections~\ref{summary}.

\section{The symmetry energy and the symmetry potential}
\label{SectionIsospin}

For completeness, we include in the following a brief introduction
to the nuclear symmetry energy and the nuclear symmetry potential
as both are closely linked to the isovector properties of asymmetric
nuclear matter. While there are extensive discussions in the
literature on the symmetry energy, relatively little attention has
been paid to the equally important symmetry potential.

\subsection{Nuclear symmetry energy}

The binding energy per nucleon of an isospin asymmetric nuclear matter
with neutron density $\rho_n$ and proton density $\rho_p$, i.e., the
EOS of the asymmetric nuclear matter can be generally expressed as a
power series in the isospin asymmetry $\delta =(\rho _{n}-\rho
_{p})/\rho$ ($\rho =\rho _{n}+\rho _{p}$ is the total nucleon density).
To the $2$nd-order in $\delta $, it can be expressed as
\begin{equation}
E(\rho ,\delta )=E_{0}(\rho )+E_{\mathrm{sym}}(\rho )\delta^{2}+O(\delta ^{4}),
\label{EOSANM}
\end{equation}%
where $E_{0}(\rho )=E(\rho ,\delta =0)$ is the binding energy per
nucleon of symmetric nuclear matter, and
\begin{eqnarray}
E_{\text{sym}}(\rho ) &=&\frac{1}{2!}\frac{\partial ^{2}E(\rho ,\delta )}
{\partial \delta ^{2}}|_{\delta =0}
\label{Esym}
\end{eqnarray}%
is the nuclear symmetry energy. The absence of odd-order terms in
$\delta $ in Eq. (\ref{EOSANM}) is due to the neutron-proton exchange
symmetry in nuclear matter when the Coulomb interaction is neglected
and the charge symmetry of nuclear forces is assumed. Empirically, the
coefficients of higher-order terms in $\delta$ are found to be negligible,
e.g., the magnitude of the coefficient of $\delta ^{4}$ term at $\rho _{0}$
is estimated to be smaller than $1$ MeV~\cite{Sie70,Sjo74,Lag81,Cai12},
compared to $\sim 30$ MeV for the coefficient of $\delta^2$ term.
Neglecting the contribution from higher-order $ \delta$ terms in
Eq. (\ref{EOSANM}) corresponds to the well-known empirical parabolic
law for the EOS of asymmetric nuclear matter, which has been verified
by all many-body theory calculations at least for densities up to
moderate values~\cite{LCK08}. The density-dependent symmetry energy
$E_{\mathrm{sym}}(\rho )$ can thus be extracted approximately from
\begin{equation}
E_{\mathrm{sym}}(\rho )\approx E(\rho ,\delta =1)-E(\rho
,\delta=0).
\label{EsymPA}
\end{equation}%
In this sense, the nuclear symmetry energy gives an estimation of
the binding energy difference between pure neutron matter and
symmetric nuclear matter. It should be noted that the possible
presence of higher-order $ \delta$ terms at supra-saturation
densities can significantly modify the proton fraction in
$\beta$-equilibrium neutron-star matter as well as the critical
density for the direct Urca process that causes the faster cooling
of neutron stars~\cite{Cai12,Zha01,Ste06}. In addition, the higher-order
$\delta $ terms have been shown to be very important in determining
the core-crust transition density and pressure in neutron
stars~\cite{Cai12,Xu09a,Xu09b}.

The binding energy per nucleon in symmetric nuclear matter $E_{0}(\rho )$
can be expanded around the saturation density $\rho _{0}$ as
\begin{eqnarray}
E_{0}(\rho ) &=&E_{0}(\rho _{0})+\frac{K_{0}}{2!}\chi ^{2}+O(\chi ^{3}),
\label{E0}
\end{eqnarray}%
where $\chi=(\rho-\rho_0)/3\rho_0$ is a dimensionless variable characterizing
the deviations of the density from $\rho _{0}$. The $E_{0}(\rho _{0})$ on the
right-hand-side of Eq. (\ref{E0}) is the binding energy per nucleon in
symmetric nuclear matter at its saturation density, and the coefficient
\begin{eqnarray}
K_{0} &=&9\rho _{0}^{2}\frac{d^{2}E_{0}(\rho )}{d\rho ^{2}}|_{\rho =\rho
_{0}},
\end{eqnarray}%
is the well-known incompressibility coefficient of symmetric nuclear
matter characterizing the curvature of $E_{0}(\rho )$ at $\rho _{0}$.
Equation (\ref{E0}) represents the parabolic approximation to the EOS of
symmetric nuclear matter. According to the definition of the saturation
density $\rho _{0}$, there is obviously no linear $\chi $ term in
Eq. (\ref{E0}).

Similarly, the nuclear symmetry energy $E_{\mathrm{sym}}(\rho )$ can also
be expanded around $\rho_0$ as
\begin{eqnarray}
E_{\mathrm{sym}}(\rho ) &=&E_{\mathrm{sym}}(\rho _{0})+L\chi
+\frac{K_{\mathrm{sym}}}{2!}\chi ^{2}+O(\chi ^{3}),
\label{EsymLKJI}
\end{eqnarray}%
where $L$ and $K_{\mathrm{sym}}$ are, respectively, the slope and
curvature parameters of the nuclear symmetry energy at $\rho _{0}$, i.e.,
\begin{eqnarray}
L &=&3\rho _{0}\frac{dE_{\mathrm{sym}}(\rho )}{d\rho }
|_{\rho =\rho _{0}},  \label{L} \\
K_{\mathrm{sym}} &=&9\rho _{0}^{2}\frac{d^{2}E_{\mathrm{sym}}(\rho )}{d^{2}\rho }|_{\rho =\rho _{0}}.
\label{Ksym}
\end{eqnarray}
The $L$ and $K_{\mathrm{sym}}$ characterize the density dependence
of the nuclear symmetry energy around the saturation density
$\rho_0$, and carry important information on both high and low
density behaviors of the nuclear symmetry energy~\cite{ChenLW11b}.

More generally, one can expand the $E_{\mathrm{sym}}(\rho )$ around an
arbitrary reference density $\rho _{r}$ as
\begin{equation}
E_{\text{sym}}(\rho )=E_{\text{sym}}({\rho _{r}})+L(\rho _{r})\chi_r+\frac{K_{\mathrm{sym}}(\rho _{r})}{2!}\chi_r ^{2}+O(\chi_r ^{3}),
\label{EsymLKr}
\end{equation}
with $\chi_r=(\rho -{\rho _{r}})/3\rho _{r}$, and the slope and
curvature parameters of the nuclear symmetry energy at $\rho _{r}$
are then defined, respectively, as
\begin{eqnarray}
L(\rho_r) &=&3\rho_r \frac{dE_{\mathrm{sym}}(\rho )}{d\rho }
|_{\rho =\rho_r }, \\
K_{\mathrm{sym}}(\rho_r) &=&9\rho_r ^{2}\frac{d^{2}E_{\mathrm{sym}}(\rho )}{d^{2}\rho }|_{\rho =\rho_r }.
\end{eqnarray}
Since different observables may probe different density regions of the
symmetry energy, the expansion in Eq. (\ref{EsymLKr}) could be very useful
in some cases~\cite{ZhangZ13}.

\subsection{Nuclear symmetry potential}

The single-nucleon potential $U_{\tau }(\rho ,\delta ,k)$ (we take
isospin index $\tau =1 $ for neutrons and $-1$ for protons if not
otherwise stated) in asymmetric nuclear matter generally depends on
the nuclear matter density $\rho $, the nuclear matter isospin
asymmetry $\delta $, and the magnitude of the nucleon momentum $k$.
Because of the isospin symmetry of nuclear interactions under the
exchange of neutrons and protons, the $U_{\tau }(\rho ,\delta ,k)$
can be expanded as a power series of $\delta $ as~\cite{XuC11,ChenR12}
\begin{eqnarray}
U_{\tau }(\rho ,\delta ,k) &=&U_{0}(\rho ,k)+\sum_{i=1,2,\cdot \cdot \cdot
}U_{{\rm sym},i}(\rho ,k)(\tau \delta )^{i}  \notag \\
&=&U_{0}(\rho ,k)+U_{{\rm sym},1}(\rho ,k)(\tau \delta )  \notag \\
&&+U_{{\rm sym},2}(\rho ,k)(\tau \delta )^{2}+\cdot \cdot \cdot ,
\label{UtauTaylor}
\end{eqnarray}%
where $U_{0}(\rho ,k)\equiv U_{n}(\rho ,0,k)=U_{p}(\rho ,0,k)$ denotes
the single-nucleon potential in symmetric nuclear matter and
$U_{{\rm sym},i}(\rho ,k)$ denotes
\begin{eqnarray}
U_{{\rm sym},i}(\rho ,k) &\equiv &\frac{1}{i!}\frac{\partial ^{i}U_{n}(\rho
,\delta ,k)}{\partial \delta ^{i}}|_{\delta =0}  \notag \\
&=&\frac{(-1)^{i}}{i!}\frac{\partial ^{i}U_{p}(\rho ,\delta ,k)}{\partial
\delta ^{i}}|_{\delta =0},
\label{defusymn}
\end{eqnarray}%
with $U_{{\rm sym},1}(\rho ,k)$ being the well-known nuclear
symmetry potential~\cite{LCK08} (which is usually denoted by $U_{\rm
sym}(\rho ,k)$) and $U_{{\rm sym},2}(\rho ,k)$ the second-order nuclear symmetry
potential. Neglecting higher-order terms (i.e., $\delta ^{2}$, $\delta ^{3}$,
$\cdot \cdot \cdot $) in Eq. (\ref{UtauTaylor}) leads to the well-known Lane
approximation~\cite{Lan62}
\begin{equation}
U_{\tau }(\rho ,\delta ,k)\approx U_{0}(\rho ,k)+U_{\rm sym}(\rho ,k)(\tau
\delta ),  \label{Lane}
\end{equation}
which has been extensively used to approximate the single-nucleon
potential in asymmetric nuclear matter. Based on the Lane
approximation, the symmetry potential $U_{\rm sym}(\rho ,k)$ can be
evaluated approximately by \cite{LCK08,LiBA04b}
\begin{equation}
U_{\rm sym}(\rho ,k)\approx \frac{U_{n}(\rho ,\delta ,k)-U_{p}(\rho ,\delta ,k)
}{2\delta }.
\label{Usym1Lane}
\end{equation}
The nuclear symmetry potential is thus related to the isovector
part of the nucleon mean-field potential in asymmetric nuclear
matter. Besides the density, the nuclear symmetry potential also
depends on the momentum or energy of a nucleon.

The nuclear symmetry potential is different from the nuclear
symmetry energy since the latter involves the integration of the
nucleon mean-field potential over the nucleon momentum. Therefore,
the nuclear symmetry energy is a thermodynamic quantity while the
nuclear symmetry potential is a dynamical quantity. On the other hand,
based on the Hugenholtz-Van Hove theorem, it has been
shown~\cite{XuC10,XuC11,ChenR12,Cai12b} that both
$E_{\text{\textrm{sym}}}({\rho })$ and $L({\rho })$ can be
completely and analytically determined from the single-nucleon
potential, especially the nuclear symmetry potential. Therefore, the
nuclear symmetry energy and the nuclear symmetry potential are
intrinsically correlated with each other. Experimentally, the
single-nucleon potential (and thus the nuclear symmetry potential)
can be obtained from the nucleon optical potential extracted from
analyzing the nucleon-nucleus scattering data, (p,n)
charge-exchange reactions between isobaric analog states, and
single-particle energy levels of bound states. These data provide
the possibility to extract information on the isospin dependence of
the nucleon optical potential, especially the energy dependence of
the nuclear symmetry potential~\cite{XuC10,LiXH13,XuC13}.

\section{Isospin- and momentum-dependent MDI interaction}
\label{SectionMDI}

In this Section, we review in detail the isospin- and
momentum-dependent MDI interaction and its implementation in
the Boltzmann-Uehling-Uhlenbeck (BUU) transport model. Also reviewed
are some results on the properties of cold asymmetric nuclear matter
obtained with the MDI interaction, such as the symmetry energy,
the symmetry potential, the nucleon effective mass, and
the in-medium NN elastic scattering cross sections.

\subsection{The MDI interaction for transport model simulations}

The nuclear single-particle potential (nuclear mean-field potential)
is a basic input in the one-body transport models for heavy ion
collisions, such as the BUU model (See, e.g., Ref.~\cite{Ber88b}
for a review). It is through the nuclear mean-field potential that
information on the nuclear matter EOS can be extracted from BUU-like
transport model simulations for heavy ion collisions. In general,
nuclear mean-field potentials are dependent on nucleon
momentum~\cite{Beh79,Dec80,Wir88,Cse92,Beh97,Beh98}.
This is evident from the observed momentum/energy dependence of
nucleon optical model potential, and can also be understood
through the exchange-term contribution from the finite-range nuclear
force. For transport model simulations for heavy ion collisions,
Gale, Bertsch, and Das Gupta (GBD)~\cite{Gal87} firstly introduced a
parametrization of momentum-dependent mean-field potential in the
BUU model. A more realistic parametrization of momentum-dependent
mean-field potential, which correctly describes the extreme
nonequilibrium situations in the early stage of heavy ion collisions,
was later introduced in the BUU model by Welke {\it et al.}~\cite{Wel88}.
It is based on a finite-range nuclear force of the Yukawa form, and
has thus been referred to as the momentum-dependent Yukawa interaction
(MDYI)~\cite{Wel88}.

In original versions of both the GBD interaction and the MDYI
interaction, the isospin dependence of nuclear mean-field potentials
was neglected. However, the well-known Lane potential~\cite{Lan62} as
observed in the momentum/energy dependent nucleon optical model
potential clearly demonstrated that the nuclear mean-field potentials
should be isospin dependent. Such potentials are expected to be
important for transport model simulations of heavy ion collisions
induced by extremely neutron(proton)-rich nuclei. Bombaci~\cite{Bom01}
extended the GBD interaction to include explicitly the isospin
dependence, resulting in the BGBD interaction. Similarly, the
inclusion of isospin dependence in the more realistic MDYI
interaction by Das {\it et al.}~\cite{Das03} leads to the so-called
MDI interaction, which will be discussed in detail in the
following.

The isospin- and momentum-dependent MDI interaction is a generalized
isospin-dependent version of the MDYI interaction with its
parameters obtained by fitting the single-particle potentials and
nuclear matter EOS predicted by the finite-range Gogny effective
interaction~\cite{Das03} using the parameter set D1~\cite{Dec80}.
In the MDI interaction, the potential energy density $V(\rho ,\delta )$
of an asymmetric nuclear matter is parameterized as follows~\cite{Das03,Che05a}:
\begin{eqnarray}
V(\rho ,\delta ) &=&\frac{A_{u}(x)\rho _{n}\rho _{p}}{\rho _{0}}+\frac{A_{l}(x)}{%
2\rho _{0}}(\rho _{n}^{2}+\rho _{p}^{2})+\frac{B}{\sigma +1}\frac{\rho
^{\sigma +1}}{\rho _{0}^{\sigma }}  \notag \\
&\times &(1-x\delta ^{2})+\frac{1}{\rho _{0}}\sum_{\tau ,\tau ^{\prime
}}C_{\tau ,\tau ^{\prime }}  \notag \\
&\times &\int \int d^{3}pd^{3}p^{\prime }\frac{f_{\tau }(\vec{r},\vec{p}%
)f_{\tau ^{\prime }}(\vec{r},\vec{p}^{\prime
})}{1+(\vec{p}-\vec{p}^{\prime })^{2}/\Lambda ^{2}}. \label{MDIV}
\end{eqnarray}%
In the mean-field approximation, Eq. (\ref{MDIV}) leads to the
following single-nucleon potential~\cite{Das03,Che05a}:
\begin{eqnarray}
U_\tau(\rho ,\delta ,\vec{p}) &=&A_{u}(x)\frac{\rho _{-\tau }}{\rho _{0}}%
+A_{l}(x)\frac{\rho _{\tau }}{\rho _{0}}  \notag \\
&+&B\left(\frac{\rho }{\rho _{0}}\right)^{\sigma }(1-x\delta ^{2})-4\tau x\frac{B}{%
\sigma +1}\frac{\rho ^{\sigma -1}}{\rho _{0}^{\sigma }}\delta \rho _{-\tau }
\notag \\
&+&\frac{2C_l}{\rho _{0}}\int d^{3}p^{\prime }\frac{f_{\tau }(%
\vec{r},\vec{p}^{\prime })}{1+(\vec{p}-\vec{p}^{\prime })^{2}/\Lambda ^{2}}
\notag \\
&+&\frac{2C_u}{\rho _{0}}\int d^{3}p^{\prime }\frac{f_{-\tau }(%
\vec{r},\vec{p}^{\prime })}{1+(\vec{p}-\vec{p}^{\prime })^{2}/\Lambda ^{2}}.
\label{MDIU}
\end{eqnarray}%
In the above, $f_{\tau }(\vec{r},\vec{p})$ is the nucleon phase-space
distribution function at coordinate $\vec{r}$ and momentum $\vec{p}$.
While the parameter $\sigma =4/3$ follows that from the Gogny interaction,
the other six parameters $A_{u}(x)$, $A_{l}(x)$, $B$, $C_{\tau ,\tau }$ $(\equiv C_l)$,
$C_{\tau ,-\tau }$ $( \equiv C_u)$, and $\Lambda $ are
obtained by fitting the momentum dependence of single-nucleon potential
to that predicted by the Gogny Hartree-Fock (and/or the
Brueckner-Hartree-Fock) calculations, the saturation properties of
symmetric nuclear matter, and the symmetry energy of $30.5$
MeV at nuclear matter saturation density $\rho _{0}=0.16$ fm$^{-3}$
\cite{Das03}. The incompressibility $K_{0}$ of cold symmetric
nuclear matter at $\rho _{0}$ is set to be $211$ MeV.

The parameters $A_{u}(x)$ and $A_{l}(x)$ depend on the parameter $x$
according to
\begin{equation}
A_{u}(x)=A_{u0}-x\frac{2B}{\sigma +1},~A_{l}(x)=A_{l0}+x\frac{2B}{\sigma +1},
\label{Ax}
\end{equation}%
with $A_{u0}=A_{u}(x=0)=-95.98$ MeV and $A_{l0}=A_{l}(x=0)=-120.57$
MeV. Varying the value of $x$ allows one to obtain different density
dependence of the nuclear symmetry energy while keeping the value of
$E_{\rm sym}(\rho_0)=30.5$ MeV and other properties of symmetric
nuclear matter unchanged~\cite{Che05a}. It thus can be adjusted to mimic
the predictions of microscopic and/or phenomenological many-body
theories on the density dependence of nuclear matter symmetry
energy. The $x$ parameter can be related to the well-known $x_3$
parameter in the density-dependent part of the Skyrme (and Gogny)
interaction~\cite{XuC10a}, i.e.,
\begin{eqnarray}
v_3(\vec{r}_1,\vec{r}_2) &=& \frac{1}{6}t_3(1+x_3 P_\sigma)
\bigg[\rho(\frac{\vec{r}_1+\vec{r}_2}{2})\bigg]^\gamma\delta(\vec{r}_1-\vec{r}_2),\nonumber\\
\label{MDI3BF}
\end{eqnarray}
by
\begin{eqnarray}
x = (1+2x_3)/3.
\end{eqnarray}
In Eq. (\ref{MDI3BF}), $\gamma$ is the density-dependence parameter
used to mimic in-medium effects of the many-body interactions
($\gamma = \sigma-1)$, and with $\gamma = 1$ it represents an
effective density-dependent two-body interaction deduced from a
three-body contact interaction in spin-saturated nuclear
matter~\cite{Vau72}. The $x_3$ is the spin (isospin)-dependence
parameter controlling the relative contributions of the density-dependent
term to the total energy in the isospin singlet ($T=0$) channel
($\propto (1+x_3)\rho^{\gamma+1}$) and triplet ($T=1$) channel
($\propto (1-x_3)\rho^{\gamma+1}$) (see, e.g., Ref.~\cite{Dec80}
for details). In particular, we note that $x_3 = 1$ ($x_3 = -1$)
means the density-dependent term only contributes to the $T=0$
($T=1$) channel. Therefore, varying $x$ from $1$ to $-1$ in
the MDI interaction can cover a large range of uncertainties due
to the spin (isospin)-dependence of the in-medium many-body
interactions. As a matter of fact, one main reason for the
rather divergent density dependence of the nuclear symmetry
energy in various Skyrme and/or Gogny Hartree-Fock calculations~\cite{Sto03}
is due to the different values of $x_3$ or $x$ parameter~\cite{XuC10b}.
On the other hand, it should be pointed out that the parameter
$x$ or $x_3$ does not affect the EOS of symmetric nuclear matter
since its contributions from $T=0$ and $T=1$ channels are exactly
canceled, i.e., $\propto(1+x_3)\rho^{\gamma+1} +
(1-x_3)\rho^{\gamma+1}=2\rho^{\gamma+1}$. Also, by the
construction in Eq.~(\ref{Ax}), the $x$ or $x_3$ parameter does not
change the symmetry energy value at $\rho_0$ from the MDI
interaction either. As to the parameter $B$, it is related to the
$t_3$ term in the Skyrme (and Gogny) effective interaction via
$B = t_3(\sigma+1)\rho_0^{\sigma}/16$, and particularly we
have $B=106.35$ MeV in the MDI interaction.

The last two terms in Eq.~(\ref{MDIU}) represent the momentum dependence
of the single-nucleon potential. The momentum dependence of the symmetry
potential comes from the different interaction strength parameters $C_l$
and $C_u$ for a nucleon of isospin $\tau $ interacting, respectively, with
like and unlike nucleons in the nuclear matter. In particular, we have
 $C_l=-11.7$ MeV and $C_u=-103.4$ MeV in the MDI interaction.

As shown in Ref.~\cite{XuJ10b}, an explicit form for an NN interaction,
which leads to a potential energy density similar to that given in
Eq.~(\ref{MDIV}) for the MDI interaction, can be obtained by assuming
that the interaction potential between two nucleons located at $\vec{r}_1$
and $\vec{r}_2$ has the following form:
\begin{eqnarray}
v(\vec{r}_1,\vec{r}_2) &=& \frac{1}{6}t_3(1+x_3 P_\sigma)
\rho^\gamma\left(\frac{\vec{r}_1+\vec{r}_2}{2}\right)
\delta(\vec{r}_1-\vec{r}_2) \notag\\
&+& (W+G P_\sigma - H P_\tau - M P_\sigma P_\tau) \frac{e^{-\mu
|\vec{r}_1-\vec{r}_2}|}{|\vec{r}_1-\vec{r}_2|}.\notag\\
\label{MDIyuk}
\end{eqnarray}
The above interaction has the same form as the Gogny interaction~\cite{Dec80,Das03}
except that the two finite-range Gaussian terms in the Gogny interaction are replaced by a single
Yukawa form. We would like to point out that a more general finite-range
effective NN interaction, which has all the four spin-parity components, i.e.,
singlet-even (SE), triplet-even (TE), singlet-odd (SO) and triplet-odd (TO) as well as
more general forms for the finite-range terms such as Yukawa, Gaussian and exponential,
has been proposed in Ref.~\cite{Beh98} and applied to study the momentum and density
dependence of the nuclear mean-field potential in symmetric nuclear matter as
well as the corresponding EOS. The same formalism has been subsequently applied to
systematically investigate neutron and proton mean-field potentials, the thermodynamic
properties of highly isospin asymmetric nuclear matter, and the properties of
neutron stars~\cite{Beh02,Beh05,Beh07,Beh11}. The eight parameters (including $x_3$) in the above NN
interaction (i.e., Eq.~(\ref{MDIyuk})) can be uniquely related to the eight parameters (including $x$)
in the MDI interaction (see the energy density functional of Eq.~(\ref{MDIV}))
through the following relations:
\begin{eqnarray}
t_3 &=& \frac{16 B}{(\sigma+1)\rho_0^\sigma},\label{t3}\\
x_3 &=& \frac{3x-1}{2},\label{x3}\\
\gamma &=& \sigma - 1,\label{gamma}\\
\mu &=& \Lambda, \label{lambda} \\
W &=& \frac{\Lambda^2}{3 \pi \rho_0} (A_1-A_2+C_l-C_u),\label{w}\\
G &=& \frac{\Lambda^2}{6 \pi \rho_0} (-A_1+A_2-4C_l+4C_u),\label{b}\\
H &=& \frac{\Lambda^2}{3 \pi \rho_0} (-2A_2-C_u),\label{h}\\
M &=& \frac{\Lambda^2}{3 \pi \rho_0} (A_2+2C_u),\label{m}
\end{eqnarray}
with $A_1=[A_l(x)+A_u(x)]/2$ and $A_2=[A_l(x)-A_u(x)]/2$.

It should be mentioned that the MDI interaction has been extensively
applied in the transport model simulations for studying isospin effects
in intermediate-energy heavy ion collisions induced by neutron-rich
nuclei~\cite{Che05a,LiBA04a,Che04,LiBA05a,LiBA05b,LiBA05c,LiBA06b,Yon06a,Yon06b,Yon07}
as well as the study on the thermal properties of asymmetric nuclear
matter~\cite{Xu07,Xu07b,Xu08} and the properties of neutron stars~\cite{Xu09a,Xu09b}.
We will highlight some results in the following.

\subsection{Cold asymmetric nuclear matter with the MDI interaction}

Although the nucleon phase space distribution function $f_{\tau
}(\vec{r},\vec{p})$ in Eq.~({\ref{MDIV}}) and Eq.~({\ref{MDIU}}) is
for nuclear matter not necessary in equilibrium, it is instructive
to examine the equilibrium case. For cold nuclear matter at zero
temperature, one has $f_{\tau}(\vec{r},\vec{p})=\frac{2}{h^{3}}\Theta (p_{F,\tau}-p)$
with $p_{F,\tau}=\hbar(3\pi^2\rho_\tau)^{1/3}$ being the Fermi
momentum of nucleons of isospin $\tau$ in asymmetric nuclear matter,
and the integrals in Eqs.~(\ref{MDIV}) and (\ref{MDIU}) can be
analytically evaluated~\cite{Das03,Che07}, leading to
\begin{eqnarray}
&\int \int& d^{3}{\vec p}d^{3}{\vec p}^{\prime }\frac{f_{\tau }(\vec{r},\vec{p}%
)f_{\tau ^{\prime }}(\vec{r},\vec{p}^{\prime
})}{1+(\vec{p}-\vec{p}^{\prime })^{2}/\Lambda ^{2}} \notag\\
&=& \frac{1}{6} \left(\frac{4 \pi}{h^3}\right)^2 \Lambda^2 \bigg\{
p_{F,\tau} p_{F,\tau^\prime} [3 (p^2_{F,\tau} + p^2_{F,\tau^\prime})
-
\Lambda^2] \notag\\
&+& 4 \Lambda \big[(p^3_{F,\tau} - p^3_{F,\tau^\prime})
\arctan\frac{p_{F,\tau} - p_{F,\tau^\prime}}{\Lambda} \notag \\
&-& (p^3_{F,\tau} + p^3_{F,\tau^\prime}) \arctan\frac{p_{F,\tau} +
p_{F,\tau^\prime}}{\Lambda}\big] \notag \\
&+& \frac{1}{4}[\Lambda^4 + 6 \Lambda^2 (p^2_{F,\tau} +
p^2_{F,\tau^\prime}) - 3 (p^2_{F,\tau} - p^2_{F,\tau^\prime})^2]
\notag\\
&\times& \ln \frac{(p_{F,\tau} + p_{F,\tau^\prime})^2 +
\Lambda^2}{(p_{F,\tau} - p_{F,\tau^\prime})^2 + \Lambda^2} \bigg\}
\end{eqnarray}
and
\begin{eqnarray}
&\int& d^{3}{\vec p}^{\prime }\frac{f_{\tau }(%
\vec{r},\vec{p}^{\prime })}{1+(\vec{p}-\vec{p}^{\prime
})^{2}/\Lambda ^{2}} \notag \\
&=& \frac{2}{h^3} \pi \Lambda^3 \bigg[\frac{p^2_{F,\tau} + \Lambda^2
- p^2}{2 p \Lambda} \ln \frac{(p + p_{F,\tau})^2 + \Lambda^2}{(p -
p_{F,\tau})^2 + \Lambda^2} \notag\\
&+& \frac{2 p_{F,\tau}}{\Lambda} - 2(\arctan\frac{p +
p_{F,\tau}}{\Lambda} - \arctan\frac{p - p_{F,\tau}}{\Lambda})\bigg].\notag\\
\end{eqnarray}

The kinetic energy contribution to the binding energy per nucleon
in cold asymmetric nuclear matter can be obtained as
\begin{eqnarray}
E_{K}(\rho,\delta) &=& \frac{1}{\rho}\int d^{3}{\vec p}
\left[\frac{p^{2}}{2
m}f_n(\vec{r},\vec{p}) + \frac{p^{2}}{2m} f_p(\vec{r},\vec{p})\right] \notag \\
&=& \frac{4 \pi}{5 m h^3 \rho} (p^5_{F,n} + p^5_{F,p}),
\end{eqnarray}
and the total binding energy per nucleon of cold asymmetric nuclear
matter then can be expressed as
\begin{equation}
E(\rho,\delta) = E_{K}(\rho,\delta) + \frac{V(\rho,\delta )}{\rho}.
\end{equation}

By setting $\rho_n=\rho_p=\rho/2$ and $p_{Fn}=p_{Fp}=p_F$,
one obtains the following EOS of cold symmetric nuclear matter:
\begin{eqnarray}
E_0(\rho) &=& \frac{8 \pi}{5 m h^3 \rho} p^5_F
+ \frac{\rho}{4 \rho_0} [A_l(x) + A_u(x)] \notag\\
&+& \frac{B}{\sigma + 1} \left(\frac{\rho}{\rho_0}\right)^\sigma +
\frac{1}{3 \rho_0 \rho}
(C_l + C_u) \left(\frac{4 \pi}{h^3}\right)^2 \Lambda^2 \notag\\
&\times& \bigg[p^2_F (6 p^2_F - \Lambda^2) - 8 \Lambda p^3_F\arctan
\frac{2 p_F}{\Lambda} \notag\\
&+& \frac{1}{4} (\Lambda^4 + 12 \Lambda^2 p^2_F)
\ln \frac{4 p^2_F + \Lambda^2}{\Lambda^2}\bigg].
\label{E0MDI}
\end{eqnarray}
It should be mentioned that, as expected, the $E_0(\rho)$ is
independent of the parameter $x$ since
$A_l(x) + A_u(x)=A_{l0} + A_{u0}=-216.55$ MeV is a
constant according to Eq.~(\ref{Ax}).

\begin{figure}[tbh]
\includegraphics[scale=0.85]{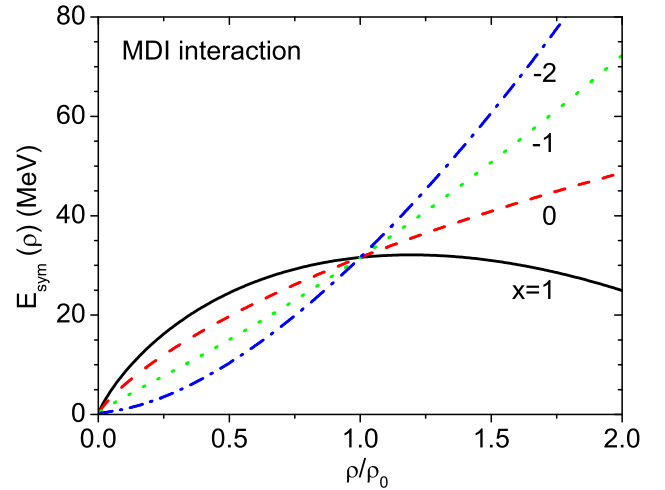}
\caption{(Color online) Symmetry energy as a function of density for
the MDI interaction with $x=1,0, -1$, and $-2$. Taken from
Ref.~\cite{Che05a}.} \label{EsymMDI}
\end{figure}

By definition, the symmetry energy can be obtained as
\begin{eqnarray}
&&E_{\rm sym}(\rho) = \frac{1}{2} \left(\frac{\partial^2 E}
{\partial \delta^2}\right)_{\delta=0} \notag\\
&=& \frac{8 \pi}{9 m h^3 \rho} p^5_f + \frac{\rho}{4 \rho_0} (A_l(x)-
A_u(x)) - \frac{B x}{\sigma + 1}
\left(\frac{\rho}{\rho_0}\right)^\sigma \notag\\
&+& \frac{C_l}{9 \rho_0 \rho} \left(\frac{4 \pi}{h^3}\right)^2
\Lambda^2 \bigg[4 p^4_F - \Lambda^2 p^2_F \ln \frac{4 p^2_F
+ \Lambda^2}{\Lambda^2}\bigg] \notag\\
&+& \frac{C_u}{9 \rho_0 \rho} \left(\frac{4 \pi}{h^3}\right)^2
\Lambda^2 \bigg[4 p^4_F - p^2_F (4 p^2_F + \Lambda^2) \ln \frac{4
p^2_F + \Lambda^2}{\Lambda^2}\bigg].\notag\\
\label{EsymMDIEq}
\end{eqnarray}
Since $A_l(x)-A_u(x)=A_{l0}-A_{u0}+4Bx/(\sigma +1)$ according to
Eq.~(\ref{Ax}), the $E_{\text{sym}}(\rho)$ depends linearly on the
parameter $x$ at a given density except $\rho_0$ where the symmetry
energy is independent of $x$ and its value is fixed by construction.
As an example, we show in Fig.~\ref{EsymMDI} the density dependence
of the symmetry energy in the MDI interaction with $x=1$, $0$, $-1$,
and $-2$. As expected, one can see that varying the value of $x$ in
the MDI interaction leads to a very broad range of the density
dependence of the nuclear symmetry energy, similar to those predicted
by various microscopic and phenomenological many-body approaches.

The single-nucleon potential for a nucleon of momentum $p$ and
isospin $\tau $ in cold asymmetric nuclear matter can be expressed
as
\begin{eqnarray}
&&U_{\tau }(\rho ,\delta ,p)=A_{u}(x)\frac{\rho _{-\tau }}{\rho _{0}}%
+A_{l}(x)\frac{\rho _{\tau }}{\rho _{0}}  \notag \\
&+&B\left(\frac{\rho }{\rho _{0}}\right)^{\sigma }(1-x\delta ^{2})-4\tau x\frac{B}{%
\sigma +1}\frac{\rho ^{\sigma -1}}{\rho _{0}^{\sigma }}\delta \rho _{-\tau }
\notag \\
&+&\frac{2C_l}{\rho _{0}}\frac{2}{h^{3}}\pi \Lambda ^{3}\Big[%
\frac{p_{F,{\tau }}^{2}+\Lambda ^{2}-p^{2}}{2p\Lambda }\ln\frac{(p+p_{F,{\tau
}})^{2}+\Lambda ^{2}}{(p-p_{F,{\tau }})^{2}+\Lambda ^{2}}  \notag \\
&+&\frac{2p_{F,{\tau }}}{\Lambda }-2\arctan\frac{p+p_{F,{\tau }}}{\Lambda }%
+2\arctan\frac{p-p_{F,{\tau }}}{\Lambda }\Big]  \notag \\
&+&\frac{2C_u}{\rho _{0}}\frac{2}{h^{3}}\pi \Lambda ^{3}\Big[%
\frac{p_{F,{-\tau }}^{2}+\Lambda ^{2}-p^{2}}{2p\Lambda }\ln\frac{%
(p+p_{F,{-\tau }})^{2}+\Lambda ^{2}}{(p-p_{F,{-\tau }})^{2}+\Lambda ^{2}}
\notag \\
&+&\frac{2p_{F,{-\tau }}}{\Lambda }-2\arctan\frac{p+p_{F,{-\tau }}}{\Lambda }%
+2\arctan\frac{p-p_{F,{-\tau }}}{\Lambda }\Big].\nonumber\\
\label{Umdi}
\end{eqnarray}%
In cold symmetric nuclear matter with $\rho_n=\rho_p=\rho/2$, the
single-nucleon potential thus is
\begin{eqnarray}
&&U_0(\rho, p)=\frac{A_l(x)+A_u(x)}{2}\frac{\rho}{\rho_0} + B\bigg(\frac{\rho }{\rho _{0}}\bigg)^{\sigma }
\notag \\
&+&\frac{2(C_{\tau ,\tau }+C_{\tau ,-\tau })}{\rho _{0}}\frac{2}{h^{3}}\pi \Lambda ^{3}\Big[%
\frac{p_{F}^{2}+\Lambda ^{2}-p^{2}}{2p\Lambda }\ln\frac{(p+p_{F})^{2}+\Lambda ^{2}}{(p-p_{F})^{2}+\Lambda ^{2}}  \notag \\
&+&\frac{2p_{F}}{\Lambda }-2\arctan\frac{p+p_{F}}{\Lambda }%
+2\arctan\frac{p-p_{F}}{\Lambda }\Big],
\label{U0mdi}
\end{eqnarray}%
and the nuclear symmetry potential by definition (i.e., Eq.~(\ref{defusymn})) is
\begin{eqnarray}
&&U_{\rm sym}(\rho,p)=\frac{A_l(x)-A_u(x)}{2}\frac{\rho}{\rho_0}-2x\frac{B}{\sigma+1}\Big(\frac{\rho}{\rho_0}\Big)^\sigma \nonumber \\
&+&\frac{2(C_{\tau,\tau}-C_{\tau,\tau'})}{\rho_0}\frac{2\pi p_F^2\Lambda^2}{3h^3p} \ln\frac{(p+p_F)^2+\Lambda^2}{(p-p_F)^2+\Lambda^2}.
\label{UsymMDIEq}
\end{eqnarray}

\begin{figure}[tbh]
\includegraphics[width=8.6cm]{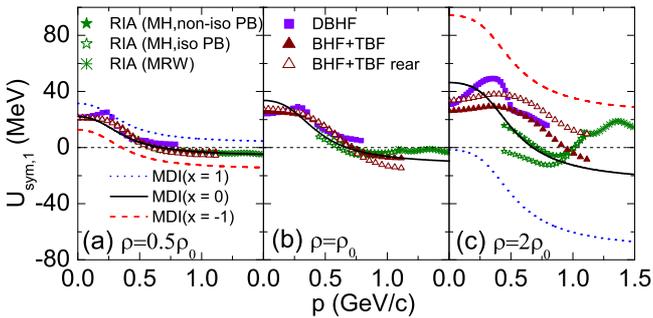}
\caption{(Color online) Momentum dependence of the symmetry
potential $U_{\rm sym,1}(\rho ,k)$ at $\rho =0.5\rho _{0}$ (a),
$\rho _{0}$ (b), and $2\rho _{0}$ (c) using the MDI
interaction with $x=-1$, $0$, and $1$. Corresponding results from
several microscopic approaches are also included for comparison (See
the text for details). Taken from Ref.~\cite{ChenR12}.}
\label{Usym1MDI}
\end{figure}

Shown in Fig.~\ref{Usym1MDI} is the momentum dependence of $U_{\rm
sym,1}(\rho ,k)$ (i.e., $U_{\rm sym}(\rho ,k)$) at $\rho =0.5\rho
_{0}$, $\rho _{0}$, and $2\rho _{0}$ in the MDI interaction
with $x=-1$, $0$, and $1$. For comparison, we also include in
Fig.~\ref{Usym1MDI} corresponding results from several microscopic
approaches, including the non-relativistic Brueckner-Hartree-Fock
(BHF) theory with and without the rearrangement contribution from
the three-body force (TBF)~\cite{Zuo06}, the relativistic
Dirac-Brueckner-Hartree-Fock (DBHF) theory~\cite{Dal05}, and the
relativistic impulse approximation (RIA)~\cite{Che05c,LiZH06b} using
the empirical NN scattering amplitude determined by Murdock and
Horowitz (MH)~\cite{Mur87} as well as by McNeil, Ray, and Wallace
(MRW)~\cite{McN83} with isospin-dependent and isospin-independent
Pauli blocking corrections. It is seen that these microscopic results
are essentially consistent with each other around and below
$\rho _{0}$ although large uncertainties still exist at higher
density $\rho =2\rho _{0}$. It is interesting to note that the
$U_{\rm sym}(\rho ,k) $ from the MDI interaction with $x=0$, which
decreases with the nucleon momentum, is in good agreement with the
results from the microscopic approaches at all densities shown in
Fig.~\ref{Usym1MDI}.

\subsection{Nucleon effective mass and in-medium nucleon-nucleon scattering
cross section with the MDI interaction}

The nucleon effective mass is an important physical quantity that
reflects the momentum dependence of the nuclear mean-field potential
in nuclear matter. In isospin asymmetric nuclear matter, the nucleon
effective mass $m^*_\tau$ is given by
\begin{equation}
\frac{m_{\tau }^{\ast }}{m_{\tau }}=\left\{ 1+\frac{m_{\tau }}{p}\frac{%
dU_{\tau }}{dp}\right\}^{-1},
\label{emass}
\end{equation}%
and it can be different for protons and nucleons. It normally depends
on the density and isospin asymmetry of the medium as well as the
momentum of the nucleon \cite{Jam89,Neg98,Fuc04}. The well-known Landau
mass which is related to the Landau parameter $f_1$ of a Fermi
liquid~\cite{Jam89,Neg98,Fuc04} corresponds to the nucleon effective
mass evaluated at the Fermi momentum $p_{\tau }=p_{F,\tau}$ in
Eq.~(\ref{emass}). A detailed discussion about different kinds of
effective masses can be found in Refs.~\cite{Jam89,Che07}. It should be
mentioned that the nucleon effective masses from the MDI interaction are
independent of the $x$ parameter because the momentum-dependent part
of the single-nucleon potential in Eq.~(\ref{MDIU}) is independent of
the parameter $x$.

\begin{figure}[tbh]
\includegraphics[height=0.35\textheight,angle=-90]{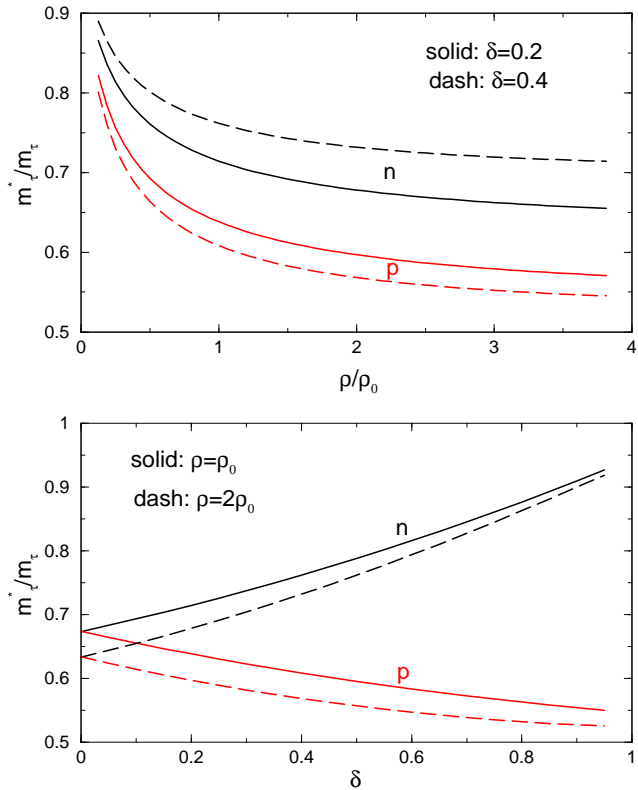}
\caption{(Color online) Neutron and proton effective masses in
asymmetric nuclear matter as functions of density (upper window) and
isospin asymmetry (lower window) from the MDI interaction.
Taken from Ref.~\cite{LiBA05c}.}
\label{EffMass}
\end{figure}

Figure~\ref{EffMass} displays the results from the MDI interaction for
the effective masses of neutrons and protons in cold asymmetric nuclear
matter at their respective Fermi surfaces (i.e., Landau mass) as
functions of density (upper window) and isospin asymmetry (lower window).
One can see that the neutron has a larger effective mass than the proton
in neutron-rich matter and the effective mass splitting between
them increases with both the density and isospin asymmetry of the
nuclear medium. It should be noted that the neutron-proton effective
mass splitting is related to the momentum dependence of the symmetry
potential~\cite{LiBA04b,XuC10}. The larger neutron effective mass
than proton effective mass in neutron-rich matter essentially
reflects the fact that the symmetry potential decreases with nucleon
momentum as shown in Fig.~\ref{Usym1MDI}, which is consistent with
calculations from many phenomenological and microscopic models, see,
e.g., Refs.~\cite{Bom91,Fuc04,Sjo76,Ma04,Zuo05,Sam05} as well as the
symmetry potential extracted from empirical isospin-dependent optical
model potentials~\cite{XuC10,LiXH13,LiXH12}.

Besides the mean-field potential, another basic component in
transport models is the NN scattering cross sections. In principle,
both nuclear mean-field potentials and NN scattering cross sections
should be determined self-consistently from the same interaction.
In practice, however, due to the complexity of the problem, the
nuclear mean-field potentials and NN scattering cross sections are
usually modeled separately in transport model simulations of heavy
ion collisions. Especially, the experimental free space NN scattering
cross sections (or with simple constant or local density-dependent
scalings) are usually used in many transport model simulations.
In the IBUU04 transport model, the option of using in-medium NN
cross sections is included by extending the effective mass scaling
model~\cite{Neg81,Pan91,LiGQ94} to isospin asymmetric matter using
the isospin- and momentum-dependent MDI interaction. In the effective
mass scaling model, the NN interaction matrix elements in the medium
are assumed to be the same as that in free-space. The NN cross
sections in the medium ($\sigma _{NN}^{\rm medium}$) thus differ
from those in free space ($\sigma_{NN}^{\rm free}$) only due to
the difference in the incoming current in the initial state
and the density of states in the final state. Since both depend on
the effective masses of a colliding nucleon pair, the NN cross
sections are thus reduced in the medium by the factor
\begin{equation}
R_{\rm medium}\equiv \sigma _{NN}^{\rm medium}/\sigma _{NN}^{\rm free}=(\mu _{NN}^{\ast
}/\mu _{NN})^{2},
\label{xmedium}
\end{equation}%
where $\mu _{NN}$ and $\mu _{NN}^{\ast }$ are, respectively, the reduced
masses of the colliding nucleon pair in free-space and in the medium.

It should be pointed out that the scaling of $\sigma _{NN}^{\rm
medium}/\sigma _{NN}^{\rm free}$ in Eq. (\ref{xmedium}) is consistent
with calculations based on the DBHF theory~\cite{Sam05b} for colliding
nucleon pairs with relative momenta less than about $240$ MeV/c at
densities less than about $2\rho _{0}$. This provides a strong support
for the effective mass scaling model in the limited density and momentum
ranges, and the scaling model to elastic NN scatterings can thus be
safely applied in the transport model simulations for heavy-ion
collisions at beam energies up to about the pion production threshold.
For heavy-ion collisions at higher energies, inelastic reaction channels
become important and in-medium effects on these channels have been a
subject of much interest~\cite{Ber88,Mao94,Cai05}. In the IBUU04 model,
the experimental free-space cross sections are, however, used for the
inelastic channels.

\begin{figure}[tbh]
\includegraphics[height=0.35\textheight,angle=-90]{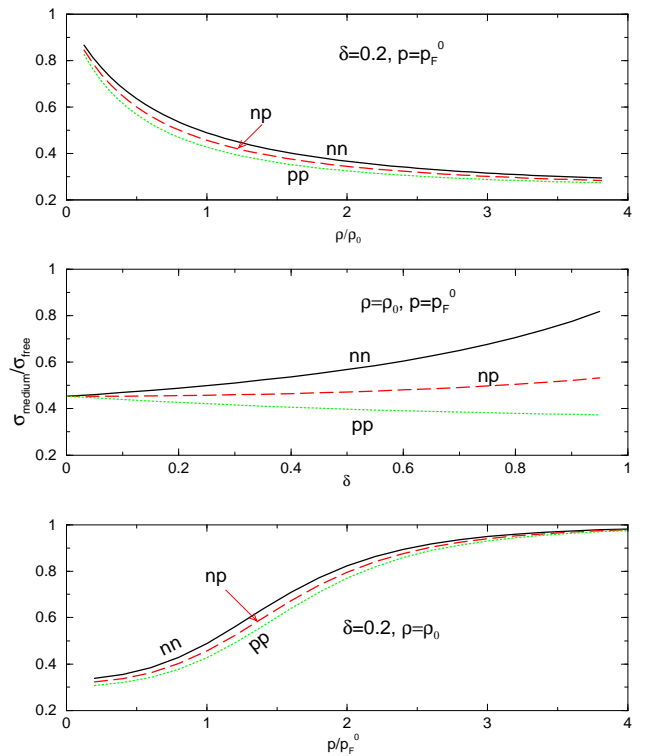}
\caption{(Color online) Ratio of NN cross sections in nuclear medium
to their free-space values as a function of density (top window),
isospin asymmetry (middle window), and momentum (bottom window).
Taken from Ref.~\cite{LiBA05c}.}
\label{medcrsc}
\end{figure}

In transport model simulations of heavy-ion collisions, the nucleon
effective masses used to obtain the in-medium NN cross sections
within the effective mass scaling model have to be evaluated
dynamically in the evolving medium created during the collisions~\cite{LiBA05c}.
As an example of a simplified case, it is instructive to see
the in-medium NN cross sections in cold asymmetric nuclear matter
in equilibrium. In this case, the nucleon effective mass can be
obtained analytically and analytical expressions of the medium
reduction factor $R_{\rm medium}$ can thus also be obtained, albeit
lengthy. Shown in Fig.~\ref{medcrsc} is the reduction factor $R_{\rm medium}$
for two colliding nucleons having the same magnitude of momentum $p$
as a function of density (upper window), isospin asymmetry (middle
window), and the nucleon momentum (bottom window). It is interesting
to see that not only the in-medium NN cross sections are reduced
compared with their free-space values, but the nn cross sections are
larger than the pp cross sections in the neutron-rich matter
although their free-space values are the same. Moreover, the difference
between the nn and pp scattering cross sections becomes larger in
more neutron-rich matter. The larger in-medium cross sections for nn
scatterings than for pp scatterings in neutron-rich matter are
completely due to the positive n-p effective mass splitting in
neutron-rich matter with the MDI interaction as shown in
Fig.~\ref{EffMass}. This feature provides a potential probe of
the n-p effective mass splitting in neutron-rich matter and can
be studied in heavy ion collisions induced by neutron-rich
nuclei. It should be noted that the in-medium NN cross sections are
also independent of the $x$ parameter in the MDI interaction and
they are solely determined by the nucleon effective masses through
the momentum dependence of the single-nucleon potential used in the
effective mass scaling model.

\section{Applications of the isospin- and momentum-dependent MDI interaction}
\label{SectionApplications}

In this Section, we highlight some applications of the MDI interaction
in studying heavy ion collisions based on transport model simulations,
the thermal properties of asymmetric nuclear matter, and the properties
of neutron stars. These studies have allowed us to extract many useful
information about the isospin and momentum dependence of the in-medium
nuclear effective interaction.

\subsection{Heavy ion collisions}

The first application of the MDI interaction was done by Li
\textit{et al.}~\cite{LiBA04a} in BUU transport model simulations of
heavy ion collisions induced by neutron-rich nuclei at intermediate
energies. They found that the symmetry potentials with and without
the momentum dependence but corresponding to the same
density-dependent symmetry energy would lead to significantly different
predictions on several symmetry energy sensitive experimental
observables. The momentum dependence of the symmetry potential is
thus very important for exploring accurately the EOS and
properties of dense neutron-rich matter. Since then, the BUU
transport model with the MDI interaction has been extensively used
for investigating the isospin effects in heavy ion collisions and for
extracting information on the density dependence of the symmetry
energy (See, e.g., Ref.~\cite{LCK08}). The BUU transport model
with the MDI interaction has also been applied to constrain the
high-density behaviors of the symmetry energy by analyzing the
FOPI data on charged pion ratio in heavy ion
collisions~\cite{Xiao09,Zha09,Zha10} as well as photon
production in those collisions~\cite{Yon08} (See, also Ref.~\cite{Xiao13}).
In this subsection, we only highlight two of these extensive
studies, i.e., the isospin diffusion/transport and t/$^3$He ratio,
with emphasis on the importance of the momentum dependence of the
mean-field potential, especially the momentum dependence of the
symmetry potential in heavy ion collisions induced by neutron-rich nuclei.

\subsubsection{Isospin diffusion/transport}

One important application~\cite{Che05a} of the MDI interaction is
the analysis of the isospin diffusion (transport) data from
NSCL-MSU~\cite{Tsa04} within the IBUU04 transport model. Experimentally,
the degree of isospin diffusion between the projectile nucleus
$A$ and the target nucleus $B$ can be studied via the physical
quantity $R_{i} $~\cite{Ram00,Tsa04} defined as
\begin{equation}
R_{i}=\frac{2X^{A+B}-X^{A+A}-X^{B+B}}{X^{A+A}-X^{B+B}},
\label{Ri}
\end{equation}%
where $X$ is an isospin-sensitive observable. One can see that the
value of $R_{i}$ is $1~(-1)$ for symmetric $A+A~(B+B)$ reaction by
construction. If isospin equilibrium is reached during the collision
due to isospin diffusion, the value of $R_{i}$ becomes about zero. In
the NSCL/MSU experiments with $A=$ $^{124}$Sn and $B=$ $^{112}$Sn at a
beam energy of $50$ MeV/nucleon and an impact parameter about $6$ fm,
the isospin asymmetry of the projectile-like residue was used as the
isospin tracer $X$~\cite{Tsa04}.

\begin{figure}[tbh]
\includegraphics[scale=0.85]{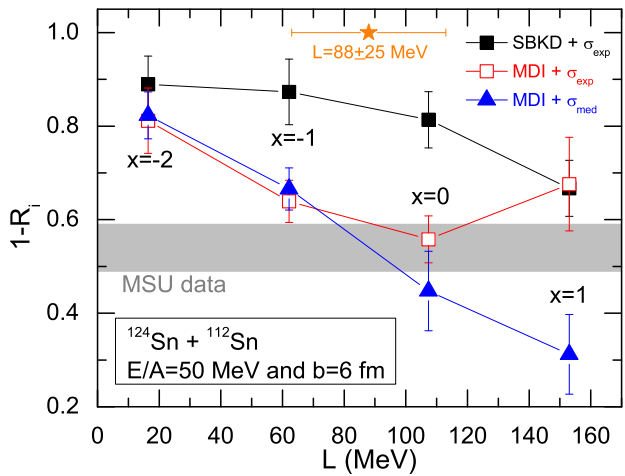}
\caption{(Color online)  Degree of the isospin diffusion $1-R_{i}$
as a function of $L$ using the SBKD interaction with the free-space
experimental nucleon-nucleon cross sections $\sigma_{\rm exp}$ (SBKD
+ $\sigma_{\rm exp}$), the MDI interaction with $\sigma_{\rm exp}$
(MDI + $\sigma_{\rm exp}$), and the MDI interaction with the
in-medium nucleon-nucleon cross sections (MDI + $\sigma_{\rm med}$).
The $x=-2$, $-1$, $0$, and $1$ indicate the different symmetry
energies in the MDI interaction and the shaded band indicates the
data from NSCL/MSU~\cite{Tsa04}. The solid star with error bar
represents $L=88\pm 25$ MeV. The calculated results are taken from
Refs.~\cite{Che05a,LiBA05c}.} \label{RiL}
\end{figure}

Shown in Fig.~\ref{RiL} is the degree of the isospin diffusion
$1-R_{i}$ as a function of the slope $L$ of the symmetry energy
at saturation density obtained from the IBUU04 transport model
with different mean-field potentials and NN cross sections, i.e.,
the MDI interaction with in-medium NN cross sections that are
consistent with the mean-field potential obtained with the MDI
interactions via the effective mass scaling model (MDI +
$\sigma_{\rm med}$), the MDI interaction with free-space
experimental NN cross sections (MDI + $\sigma_{\rm exp}$), and
the momentum-independent soft Bertsch-Kruse-Das Gupta (SBKD)~\cite{Ber84}
mean-field potential with free-space experimental NN cross sections.
For the momentum-independent SBKD mean-field potential, the
momentum-independent symmetry potential
obtained from the same density-dependent symmetry energy with
different $x$ values has been used~\cite{Che05a}. The shaded band
in Fig.~\ref{RiL} indicates the data from NSCL/MSU~\cite{Tsa04}.
It is seen from Fig.~\ref{RiL} that for the SBKD interaction without
momentum dependence, the isospin diffusion decreases monotonically
(i.e., increasing value for $R_i$) with increasing $L$. The isospin
diffusion is reduced when the isospin- and momentum-dependent MDI
interaction (with $\sigma_{\rm exp}$) is used because the
momentum dependence weakens the strength of symmetry potential
except for $x=-2$. As seen in Fig. 3 of Ref.~\cite{Che05a}, the
symmetry potential in the MDI interaction has the smallest strength
for $x=-1$ as it is close to zero at $k\approx 1.5$ fm$^{-1}$ and
$\rho /\rho _{0}\approx 0.5$, and increases again with further
stiffening of the symmetry energy, e.g., $x=-2$, when it becomes
largely negative at all momenta and densities. The MDI interaction
with $x=-1$ thus gives the smallest degree of isospin diffusion
among the interactions considered and reproduces the MSU data.
If the isoscalar part of the SBKD potential is replaced with the
momentum-dependent (but isospin-independent) MDYI potential,
which has a similar $K_{0}$ as those for the MDI and SBKD interactions,
the resulting $R_{i}=0.37\pm 0.07$ is close to that obtained in the
MDI interaction ($R_{i}=0.44\pm 0.05$) with $x=-1$, implying that the
momentum dependence of the symmetry potential introduced in the MDI
interaction leads to about $16\%$ variation for the isospin
diffusion.

Furthermore, one can see from Fig.~\ref{RiL} that the difference in
$1-R_{i}$ obtained with the free-space (MDI + $\sigma_{\rm exp}$)
and the in-medium (MDI + $\sigma_{\rm med}$) NN cross sections
using the same MDI interaction is about the same for $x=-2$ and $x=-1$,
but becomes especially large at $x=0$ and $x=1$. As pointed out in
Ref.~\cite{LiBA05c}, these results can be understood by considering
contributions from the symmetry potential and the np scatterings.
Schematically, the mean-field contribution is proportional to the
product of the isospin asymmetric force $F_{\rm np}$ and the inverse of
the np scattering cross section $\sigma_{np}$~\cite{Shi03}. While the
collisional contribution is proportional to $\sigma _{np}$, the
overall effect of NN cross sections on isospin diffusion
is a result of a complicated combination of the effects due to both
the nuclear mean field and NN scatterings. Generally speaking, while
the symmetry potential effects on the isospin diffusion become
weaker when the NN cross sections are larger, the symmetry potential
effects would show up more clearly if NN cross sections are smaller.

From comparison of the theoretical results with MDI + $\sigma_{\rm
med}$ to the data, one can extract a value of $L=88\pm 25$ MeV,
shown by the solid star with error bar in Fig.~\ref{RiL}. This
value (and $E_{\text{\textrm{sym}}}({\rho _{0}}) = 31.6$ MeV) has
been obtained with the parabolic approximation (i.e., Eq~(\ref{EsymPA}))
for the symmetry energy. Without using the parabolic approximation, the
constraint changes slightly to $L = 86 \pm 25$ MeV with
$E_{\text{\textrm{sym}}}({\rho _{0}}) = 30.5$ MeV. This nuclear
symmetry energy is significantly softer than the prediction by
transport model simulation with a momentum-independent
interaction~\cite{Tsa04} but is in agreement with a number of
microscopic theoretical calculations. These results indicate that
the isospin diffusion in heavy ion collisions indeed provides a
sensitive probe of the isospin- and momentum-dependent nuclear
effective interaction and corresponding in-medium NN scattering
cross sections.

\subsubsection{t/$^3$He ratio}

While the neutron/proton ratios of preequilibrium
nucleons~\cite{LiBA97a} and squeezed-out nucleons~\cite{Yon07} in
heavy ion collisions induced by neutron-rich nuclei have been shown
to be sensitive probes of the density dependence of the symmetry
energy, they suffer from some practical difficulties since it is
hard to measure neutrons accurately in experiments. On the other
hand, for light charged clusters such as deutron (d), triton (t), and
$^3$He, their yields and ratios have also been shown to be sensitive
to the density dependence of the symmetry energy~\cite{Che03b,Che04},
and compared to neutrons they can be measured much easier in
experiments.

Light cluster production has been extensively investigated in
experiments involving heavy ion collisions at all energies (see,
e.g., Ref.~\cite{Hod03} for a review). A popular model for
describing the production of light clusters in these collisions
is the coalescence model, e.g., see Ref.~\cite{Cse86} for a
theoretical review, which has been used at both
intermediate~\cite{Gyu83,Aich87,Koch90,Indra00} and high
energies~\cite{Mat95,Nag96}. In this model, the cluster production
probability is determined by the overlap of its Wigner phase-space
density with the phase-space distribution of nucleons at
freeze out in a heavy ion collision. Explicitly, the multiplicity
of an $M$-nucleon cluster is given by~\cite{Mat95}
\begin{eqnarray}
N_{M}=G\int d\mathbf{r}_{i_{1}}d\mathbf{q}_{i_{1}}\cdots d\mathbf{r}%
_{i_{M-1}}d\mathbf{q}_{i_{M-1}} \notag \\
\langle \underset{i_{1}>i_{2}>...>i_{M}}{\sum
}\rho _{i}^{W}(\mathbf{r}_{i_{1}},\mathbf{q}_{i_{1}}\cdots \mathbf{r}%
_{i_{M-1}},\mathbf{q}_{i_{M-1}})\rangle .
\end{eqnarray}%
In the above, $G$ is the spin-isospin statistical factor for the
cluster~\cite{Pol99}; $\mathbf{r}_{i_{1}}$, $\cdots$, $\mathbf{r}_{i_{M-1}}$
and $\mathbf{q}_{i_{1}}$, $\cdots$, $\mathbf{q}_{i_{M-1}}$ are,
respectively, the $M-1$ relative coordinates and momenta taken at equal
time in the rest frame of the $M$ nucleons; $\rho _{i}^{W}$ is the Wigner
phase-space density of the $M$-nucleon cluster; and
$\langle \cdots \rangle $ means event averaging. Details about such
calculation can be found in Ref.~\cite{Che03b}.

\begin{figure}[th]
\includegraphics[scale=0.95]{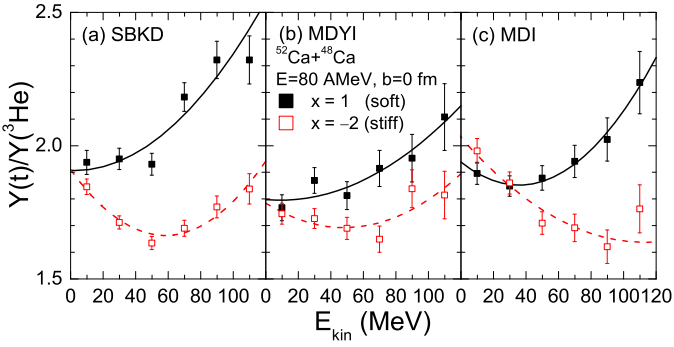}
\caption{(Color online) The cluster kinetic energy dependence of
the t/$^{3}$He ratio for interactions SBKD (a), MDYI (b), and
MDI (c) with the soft (solid squares) and stiff (open squares)
symmetry energies. The lines are drawn to guide eyes. Taken from
Ref.~\cite{Che04}.}
\label{RtHe3}
\end{figure}

The isobaric yield ratio t/$^{3}$He is less dependent on the
isoscalar properties of the nuclear mean-field potential and
also less affected by other effects, such as the feedback from
heavy fragment evaporation and from excited triton and
$^{3}$He states produced in heavy ion collisions, and thus
provides a good probe of the density dependence of the
symmetry energy~\cite{Che03b,Che04}. Shown in Fig.~\ref{RtHe3}
is the t/$^{3}$He ratio as a function of cluster kinetic energy
(in the center-of-mass system) from central collisions of
$^{52}$Ca + $^{48}$Ca at $E=80$ MeV/nucleon by using the SBKD
(left panel), MDYI (middle panel), and MDI (right panel)
interactions with the soft (solid squares) and stiff (open squares)
symmetry energies. Here, the soft (stiff) symmetry energy from
the SBKD or MDYI interaction corresponds to the same density-dependent
symmetry energy obtained from the MDI interaction
with $x=1 (-2)$. For all cases, the free-space experimental NN
cross sections are used. It is seen from Fig.~\ref{RtHe3} that
for all nuclear interactions the ratio t/$^{3}$He exhibits very
different energy dependence for different symmetry energies.
While the t/$^{3}$He ratio decreases and/or increase weakly
with kinetic energy for the stiff symmetry energy, it increases
with kinetic energy for the soft symmetry energy. For both stiff
and soft symmetry energies, the value of t/$^{3}$He ratio is
larger than that of the neutron to proton ratio of the whole
reaction system, i.e., N/Z$=1.5$, consistent with
results from both experiments and the statistical model
calculations for other reaction systems and incident
energies~\cite{Hagel00,Cibor00,Sobotka01,Vesel01,Chomaz99}.

It is interesting to note that for the isospin- and
momentum-dependent MDI interaction, the t/$^{3}$He ratio
displays very different energy dependence for the soft and
stiff symmetry energies, although their individual yields are not
so sensitive to the density dependence of symmetry
energy~\cite{Che03b,Che04}. This is related to the different
momentum-dependent symmetry potentials at different densities
in the MDI interaction. Also, one can see from comparing
Fig.~\ref{RtHe3} (a) and (b) that the momentum dependence
of the isoscalar potential has obvious effects on the energy
dependence of the t/$^{3}$He ratio. These results indicate that
the energy dependence of the t/$^{3}$He ratio in heavy ion
collisions induced by neutron-rich nuclei indeed provides a
sensitive probe of the isospin- and momentum-dependent
single-nucleon potential in asymmetric nuclear matter.

\subsection{Thermal properties of asymmetric nuclear matter}

\subsubsection{Temperature dependence of symmetry energy and symmetry free energy}

While the exact knowledge on the symmetry energy at zero temperature
is important for understanding ground state properties of exotic nuclei
and properties of old neutron stars at $\beta $-equilibrium,
the symmetry energy or symmetry free energy at finite temperature
is important for determining the liquid-gas phase transition of
asymmetric nuclear matter, the dynamical evolution of compact stars,
and the explosion mechanisms of
supernova~\cite{Xu07,Xu08,Che01,Zuo03,LiBA06,Mou07,De12,Don94,Dea95}.

For an asymmetric nuclear matter at thermal equilibrium with a finite
temperature $T$, the nucleon phase space distribution function becomes
the Fermi-Dirac distribution
\begin{equation}
f_{\tau
}(\vec{r},\vec{p})=\frac{2}{h^{3}}\frac{1}{\exp\left(\frac{p^{2}/
2m_{_{\tau }}+U_{\tau }-\mu _{\tau }}{T}\right)+1}, \label{f}
\end{equation}%
where $\mu _{\tau }$ is the chemical potential of proton or neutron and
can be obtained from%
\begin{equation}
\rho _{\tau }=\int f_{\tau }(\vec{r},\vec{p})d^{3}{\vec p}.
\end{equation}%
For fixed density $\rho $, isospin asymmetry $\delta $, and temperature
$T$, the chemical potential $\mu _{\tau }$ and the nucleon distribution
function $f_{\tau }(\vec{r},\vec{p})$ can be determined
numerically by a self-consistency iteration recipe~\cite{Gal90,Xu07}.
The energy per nucleon $E(\rho ,T,\delta )$ can then be obtained as
\begin{equation}
E(\rho ,T,\delta )=\frac{1}{\rho }\left[ V(\rho ,T,\delta )+{\sum_{\tau }}%
\int d^{3}{\vec p}\frac{p^{2}}{2m_{\tau }}f_{\tau
}(\vec{r},\vec{p})\right], \label{E}
\end{equation}%
while the entropy per nucleon $S_{\tau }(\rho ,T,\delta )$ is
\begin{equation}
S_{\tau }(\rho ,T,\delta )=-\frac{8\pi }{{\rho }h^{3}}\int_{0}^{\infty
}p^{2}[n_{\tau }\ln n_{\tau }+(1-n_{\tau })\ln (1-n_{\tau })]dp  \label{S}
\end{equation}%
with the occupation probability%
\begin{equation}
n_{\tau }=\frac{1}{\exp \left(\frac{p^{2}/2m_{_{\tau }}+U_{\tau
}-\mu _{\tau }}{T}\right)+1}.
\end{equation}%
Correspondingly, the free energy per nucleon $F(\rho ,T,\delta )$ and
the pressure $P(\rho ,T,\delta )$ of the thermal equilibrium asymmetric
nuclear matter can be obtained from the thermodynamic relations,
\begin{eqnarray}
F(\rho ,T,\delta ) &=& E(\rho ,T,\delta )-T{\sum_{\tau }}S_{\tau }(\rho
,T,\delta ), \label{F} \\
P(\rho ,T,\delta ) &=& \sum_{\tau }\mu _{\tau }\rho _{\tau } - F(\rho ,T,\delta )  \rho.
\label{P}
\end{eqnarray}

As in the situation of zero temperature, phenomenological and microscopic
studies~\cite{Che01,Zuo03} have shown that the EOS of hot neutron-rich
matter at density $\rho $, isospin asymmetry $\delta $, and temperature $T$,
can also be written as a parabolic function of $\delta $, i.e.,
\begin{equation}
E(\rho ,T,\delta )=E(\rho ,T,\delta =0)+E_{\rm sym}(\rho ,T)\delta ^{2}+\mathcal{%
O}(\delta ^{4}).  \label{eos}
\end{equation}%
The density- and temperature-dependent symmetry energy $E_{\rm sym}(\rho ,T)$ for
hot neutron-rich matter can thus be extracted from
$E_{\rm sym}(\rho ,T)\approx E(\rho ,T,\delta =1)-E(\rho,T,\delta =0)$.
Similarly, one can define the density and temperature dependent symmetry free
energy $F_{\rm sym}(\rho ,T)$ by the following parabolic approximation to the
free energy per nucleon:
\begin{equation}
F(\rho ,T,\delta )=F(\rho ,T,\delta =0)+F_{\rm sym}(\rho ,T)\delta ^{2}+\mathcal{%
O}(\delta ^{4}).  \label{eosF}
\end{equation}%
The above parabolic approximation to the free energy per nucleon has been shown
to be a good approximation~\cite{Xu07,Xu08}, and this leads to
$F_{\rm sym}(\rho,T)\approx F(\rho ,T,\delta=1)-F(\rho ,T,\delta =0)$.

\begin{figure}[tbh]
\includegraphics[scale=0.9]{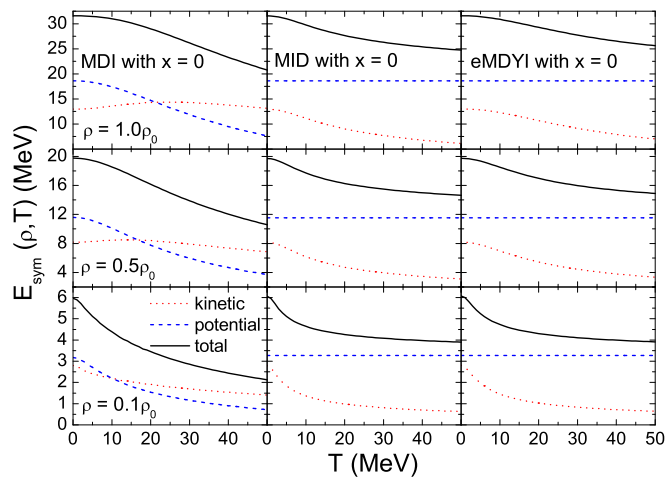}
\caption{(Color online) Total symmetry energy and its kinetic and potential
parts as functions of temperature in MDI, MID, and eMDYI interactions
with $x=0$ at $\rho=0.1\rho _{0}$, $0.5\rho _{0}$, and $1.0\rho_{0}$. Taken
from Ref.~\cite{Xu08}.}
\label{EsymTx0}
\end{figure}

Shown in Fig.~\ref{EsymTx0} is the temperature dependence of the
symmetry energy $E_{\rm sym}(\rho ,T)$ as well as its potential and
kinetic energy parts in the MDI, MID, and eMDYI interactions
with $x=0$ at $\rho =1.0\rho _{0}$, $0.5\rho _{0}$, and
$0.1\rho _{0}$. The results are similar for other $x$ values such as
$x=-1$. We note that the MID interaction corresponds to the
momentum-independent BKD energy density functional while the eMDYI
interaction corresponds to the isoscalar momentum-dependent MDYI
energy density functional but with their parameters refitted to the
density-dependent EOS from the MDI interaction (See
Ref.~\cite{Xu07,Xu07b,Xu08} for details). In the eMDYI interaction,
the resulting single-nucleon potential is momentum dependent
but its momentum dependence is isospin independent. Comparing
the eMDYI results with those from the MDI interaction, one can extract
information about the effects of the momentum dependence
of the symmetry potential, while the effects of the momentum
dependence of the isoscalar part of the single-nucleon potential
can be investigated by comparing the eMDYI results with those from
the MID interaction.

For the MDI interaction, one can see from Fig.~\ref{EsymTx0} that
both the total symmetry energy $E_{\rm sym}(\rho ,T)$ and its
potential energy part decrease with increasing temperature at
all three considered densities. The kinetic contribution, on the other
hand, increases slightly with increasing temperature at low
temperatures and then decreases with increasing temperature at
high temperatures for $\rho =1.0\rho _{0}$ and $0.5\rho _{0}$,
while it decreases monotonically for $\rho =0.1\rho _{0}$. These
features are uniquely determined by the isospin and momentum
dependence in the MDI interaction. For MID and eMDYI interactions,
the kinetic part of the total symmetry energy decreases with
increasing temperature at all considered densities, while the
potential part is independent of temperature and has the
same value for the MID and eMDYI interactions. These results
indicate that the temperature dependence of the total symmetry
energy is due to both the potential and kinetic contributions for
the MDI interaction, but it is only due to the kinetic contribution
for the MID and eMDYI interactions.

The decrement of the kinetic energy part of the symmetry energy
with temperature at very low densities is consistent with the results
from the free Fermi gas model at high temperatures and/or very low
densities~\cite{LiBA06,Lee01,Mek05,Mou07}. Also, the temperature
dependence of the total symmetry energy $E_{\rm sym}(\rho ,T)$ is
quite similar for all three interactions except that the MDI
interaction displays a slightly stronger temperature dependence at
higher temperatures. This is due to the fact that the nucleon
phase-space distribution function varies self-consistently
whether the single-nucleon potential is momentum
dependent or not. As shown in Ref.~\cite{Mou07}, both the potential
and kinetic parts of the symmetry energy $E_{\rm sym}(\rho ,T)$
also decrease with temperature for all considered densities when
using the isospin- and momentum-dependent BGBD interaction. The
different temperature dependence of the potential and kinetic parts
of the symmetry energy between the MDI and BGBD interaction is due to
the fact that the MDI and BGBD interactions have different forms for
the energy density functional with the MDI interaction having a more
complicated momentum dependence in the single-nucleon potential as
mentioned earlier. This feature implies that the temperature dependence
of the potential and kinetic parts of the symmetry energy depends on
the isospin and momentum dependence of the in-medium nuclear effective
interactions.

Experimentally, it is still a big challenge to determine the
temperature dependence of the symmetry energy or symmetry free energy.
It has been found both experimentally and theoretically that
in many types of reactions the yield ratio $R_{21}(N,Z)$ of a fragment
with proton number $Z$ and neutron number $N$ from two reactions
reaching about the same temperature $T$ respects an exponential
relationship $R_{21}(N,Z)$ $\propto$ $\exp (\alpha N)$ \cite%
{betty01,shetty,sjy,shetty06,indra,wolfgang,kowalski,tsang01,botvina,ono,dorso,ma}%
. In particular, it has been shown in several statistical and
dynamical models under some assumptions~\cite{tsang01,botvina,ono} that
the scaling coefficient $\alpha $ can be related to the symmetry energy
$C_{\rm sym}(\rho ,T)$ via
\begin{equation}
\alpha =\frac{4C_{\rm sym}(\rho ,T)}{T}\bigtriangleup \lbrack (Z/A)^{2}],
\label{scaling}
\end{equation}%
where $\bigtriangleup \lbrack (Z/A)^{2}]\equiv
(Z_{1}/A_{1})^{2}-(Z_{2}/A_{2})^{2}$ is the $(Z/A)^{2} $ difference
between the two fragmenting sources created in the two reactions. As
pointed out in Ref.~\cite{LiBA06}, because of the different assumptions
used in the various derivations, the validity of Eq. (\ref{scaling}) is
still under debate as to whether and when the $C_{\rm sym}$ is
actually the symmetry energy or the symmetry free energy. In addition,
the physical interpretation for $C_{\rm sym}(\rho ,T)$ is also
controversial, sometimes even contradictory, in the literature. The main
issue is whether the $C_{\rm sym}$ measures the symmetry (free) energy
of the fragmenting source or that of the fragments formed at freeze-out.
This ambiguity also comes from the fact that the derivation of
Eq. (\ref{scaling}) is not unique. In particular, within the grand
canonical statistical model for multifragmentation~\cite{tsang01,botvina},
the $C_{\rm sym}$ is identified as the symmetry energy of primary fragments.
While within the sequential Weisskopf model in the grand canonical
limit~\cite{tsang01}, it refers to the symmetry energy of the emission
source. Following the arguments in Ref.~\cite{LiBA06}, we assume here that
the $C_{\rm sym}$ reflects the symmetry energy of \emph{bulk nuclear matter}
for the emission source.

The temperature dependence of the symmetry energy has been studied based on
a simplified degenerate Fermi gas model~\cite{LiBA06}, and it was shown that
the experimentally observed decrease of the symmetry energy with the
increasing excitation energy or centrality in isotopic scaling analyses
of heavy ion collisions can be well understood analytically within the
degenerate Fermi gas model. In particular, it was found that the
evolution of the symmetry energy extracted from the isotopic scaling
analysis is mainly due to the variation in the freeze-out density
rather than the temperature when fragments are emitted in the
reactions.

\begin{figure}[tbh]
\includegraphics[scale=0.85]{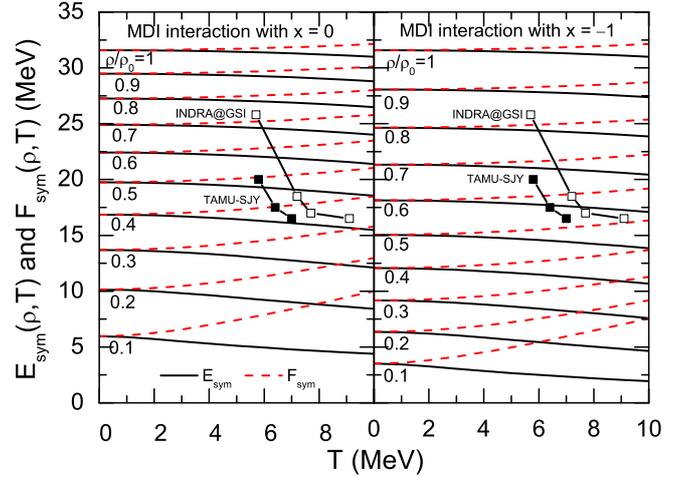}
\caption{(Color online) Temperature dependence of the
symmetry energy (solid lines) and symmetry free energy (dashed lines) using
MDI interaction with $x=0$ (left panel) and $-1$ (right panel) at different
densities from $0.1\rho _{0}$ to $\rho _{0}$. The experimental data from
Texas A$\&$M University (solid squares) and the INDRA-ALADIN collaboration
at GSI (open squares) are included for comparison. Taken from Ref.~\cite{Xu07}.}
\label{EFsymTData}
\end{figure}

Shown in Fig.~\ref{EFsymTData} is the symmetry energy $E_{\rm sym}(\rho ,T)$
and symmetry free energy $F_{\rm sym}(\rho ,T)$ as functions of temperature
using the more realistic MDI interaction with $x=0$ and $-1$ at different
densities from $0.1\rho_{0} $ to $\rho _{0}$. One can see that while the
symmetry energy $E_{\rm sym}(\rho ,T)$ deceases slightly with increasing
temperature at a given density, the symmetry free energy $F_{\rm
sym}(\rho ,T)$ increases instead. Around the saturation density
$\rho _{0}$, the difference between the symmetry energy $E_{\rm sym}(\rho ,T)$
and the symmetry free energy $F_{\rm sym}(\rho ,T)$ is quite small compared
with their values at $T=0$ MeV. This feature supports the assumption on
identifying $C_{\rm sym}(\rho ,T)$ to $E_{\rm sym}(\rho ,T)$ at lower
temperatures and not so low densities~\cite{shetty,sjy,shetty06}. On the
other hand, the symmetry free energy $F_{\rm sym}(\rho ,T)$ at low densities
displays a stronger temperature dependence, and it is significantly larger
than the symmetry energy $E_{\rm sym}(\rho ,T)$ at moderate and high
temperatures. This can be understood from the fact that the entropy
contribution to the symmetry free energy $F_{\rm sym}(\rho ,T)$ becomes
important at low densities. We would like to point out that the entropy and
thus the symmetry free energy at low densities are affected strongly by
the formation of clusters~\cite{kowalski,horowitz,Nat10,Typ13}, which are
not considered here in the mean-field calculations.

\subsubsection{Liquid-gas phase transition of asymmetric nuclear matter}

The feature of short-range repulsion and long-range attraction in the
nucleon-nucleon interaction has led to the expectation that
the liquid-gas (LG) phase transition should also exist in nuclear matter.
Since the early work, see, e.g., Refs.~\cite{lamb78,fin82,ber83,jaqaman83},
a lot of studies have been devoted to investigating the properties of the
nuclear LG phase transition both experimentally and theoretically during
the past three decades (see, e.g., Refs.~\cite{chomaz,das,wci} for a recent
review). Most of these investigations focused on studying features of the
LG phase transition in symmetric nuclear matter. Theoretically, new
features of the LG phase transition in isospin asymmetric nuclear matter
are expected. In particular, because of the two conserved charges of
baryon number and isospin due to the two components of protons and
neutrons in an asymmetric nuclear matter, the LG phase transition has
been suggested to be of second order~\cite{Mul95}. Since the isovector
nuclear interaction and the density dependence of the nuclear symmetry
energy play a central role in understanding the thermal properties of
asymmetric nuclear matter~\cite{wci,LiBA98,LiBA01b}, it is therefore
of great interest to investigate how the isospin and momentum dependence
of the nuclear effective interactions affect the LG phase transition in
asymmetric nuclear matter.

According to the Gibbs conditions for the phase coexistence of the
LG phase transition, the two-phase coexistence equations in hot
asymmetric nuclear matter can be expressed as
\begin{eqnarray}
P^{L}(T,\rho ^{L},\delta ^{L}) &=&P^{G}(T,\rho ^{G},\delta ^{G}),
\label{coexistenceP} \\
\mu _{n}^{L}(T,\rho ^{L},\delta ^{L}) &=&\mu _{n}^{G}(T,\rho ^{G},\delta
^{G}),  \label{coexistencemuN} \\
\mu _{p}^{L}(T,\rho ^{L},\delta ^{L}) &=&\mu _{p}^{G}(T,\rho ^{G},\delta
^{G}),  \label{coexistencemuP}
\end{eqnarray}%
where $L$ ($G$) stands for the liquid (gas) phase. The Gibbs phase
equilibrium conditions require equal pressures and
chemical potentials for two phases with different densities and
isospin asymmetries. For a fixed pressure, the two solutions
form the edges of a rectangle in the neutron and proton chemical
potential isobars as a function of isospin asymmetry $\delta $, and
this can be found through the geometrical construction
method~\cite{Mul95,Su00,Xu07b}. For each interaction, the two
different values of $\delta $ correspond to two different phases
with different densities, and the higher density phase (with
smaller $\delta $ value) defines the liquid phase while the lower
density phase (with larger $\delta $ value) defines the gas phase.
The binodal surface then can be obtained with all such pairs of
$\delta (T,P)$ and $\delta ^{\prime }(T,P)$.

\begin{figure}[tbh]
\includegraphics[scale=0.88]{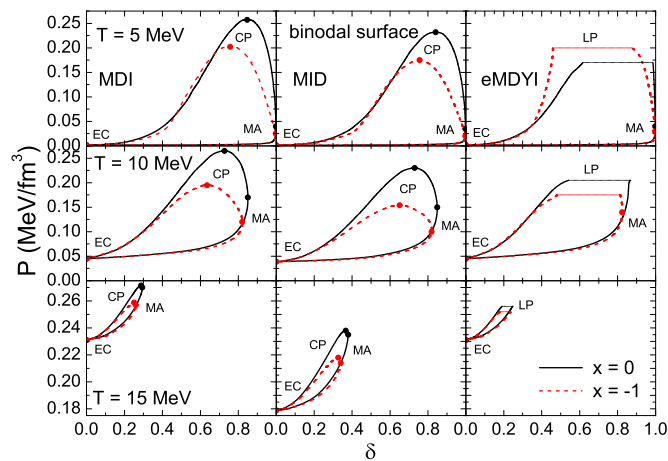}
\caption{(Color online) The section of binodal surface at $T=5$,
$10$, and $15$ MeV in the $P$-$\delta $ plane from the MDI, MID,
and eMDYI interactions with $x=0$ and $x=-1$.
Taken from Ref.~\cite{Xu08}.}
\label{Pdelta}
\end{figure}

Shown in Fig.~\ref{Pdelta} is the section of binodal surface at
$T=5$, $10$, and $15$ MeV in the $P$-$\delta $ plane from the MDI,
MID, and eMDYI interactions with $x=0$ and $x=-1$. The critical
pressure (CP) and the limiting pressure (LP) as well as the equal
concentration (EC) and the maximal asymmetry (MA) points are also
indicated in Fig.~\ref{Pdelta}. One can see that for the MDI and MID
interactions the binodal surface has a critical pressure, while
for the eMDYI interaction the binodal surface is cut off by a
limit pressure which is due to the specific momentum dependence
in the eMDYI interaction as discussed in Ref.~\cite{Xu07b}. There
is no phase-coexistence region above the critical pressure or
below the pressure of equal concentration point. Moreover,
the right side of the binodal surface is the region of gas phase
while the left side is the region of liquid phase, and the inside
of the surface corresponds to the phase-coexistence region.

It is seen from Fig.~\ref{Pdelta} that the stiffness of the symmetry
energy significantly affects the critical pressure, with a softer
symmetry energy ($x=0$) giving a higher critical pressure and a
larger area of phase-coexistence region. For the limit pressure from
the eMDYI interaction, this holds true at $T=10$ MeV and $T=15$ MeV,
but the opposite result is observed at $T=5$ MeV. Comparing the results
from the MDI interaction to those from the MID interactions, one can
see that the isospin and momentum dependence of the nuclear effective
interaction in the MDI interaction seems to increase the critical
pressure by a larger amount. In addition, at $T=5$ MeV and $T=10$ MeV,
the area of phase-coexistence region for the MID interaction is
smaller than that for the MDI interaction, but the opposite result
is obtained at $T=15$ MeV.

\subsubsection{Differential isospin fractionation}

As pointed out above, the lower density gas phase is more
neutron-rich than the coexisting liquid phase. This feature leads
to the so-called isospin fractionation (IsoF) phenomenon that has
been observed in heavy ion reaction experiments, see, e.g.,
Ref.~\cite{Xu00}. The nonequal partition of the system's isospin
asymmetry, i.e., the IsoF, has been found to be a general phenomenon
in essentially all thermodynamical models and transport model
simulations of heavy ion collisions (see, e.g.,
Refs.~\cite{LCK08,chomaz,das,Bar05} for reviews ).

In almost all existing theoretical and experimental studies on
IsoF in the literature, only the ratio between all neutrons and
protons in the liquid or the gas phase was considered,
which is normally referred to as the integrated IsoF.
On the other hand, it has been shown~\cite{LiBA07} that completely
new and very interesting physics can be revealed from the differential
IsoF that takes consideration of the nucleon momentum dependence of
the neutron/proton ratio in the liquid or the gas phase. For energetic
nucleons, with their differential IsoF very sensitive to the momentum
dependence of the symmetry potential, the nucleon phase-space
distribution function $f_{\tau }$ can be well approximated by
the Boltzmann distribution. For these nucleons in either the
liquid ($L$) or gas ($G$) phase, the neutron/proton ratio can
be expressed as
\begin{equation}
(n/p)_{L/G}=\exp [-(E_{n}^{L/G}-E_{p}^{L/G}-\mu _{n}^{L/G}+\mu
_{p}^{L/G})/T].
\end{equation}%
The energy difference of neutrons and protons having the same
kinetic energy and mass, i.e.,
\begin{equation}
E_{n}^{L/G}-E_{p}^{L/G}=U_{n}^{L/G}-U_{p}^{L/G}\approx 2\delta _{L/G}\cdot
U_{\rm sym}(p,\rho _{L/G})
\end{equation}%
is directly linked to the symmetry potential $U_{\rm sym}$.
Because of the chemical equilibrium conditions given in Eqs.
(\ref{coexistencemuN}) and (\ref{coexistencemuP}), the chemical
potentials cancel out exactly in the double neutron/proton ratio and
this leads to
\begin{equation}
\frac{(n/p)_{G}}{(n/p)_{L}}(p)=\exp [-2(\delta _{G}\cdot U_{\rm sym}(p,\rho
_{G})-\delta _{L}\cdot U_{\rm sym}(p,\rho _{L}))/T].
\end{equation}%
This general expression clearly indicates that the differential IsoF for
energetic nucleons can carry direct information on the momentum dependence
of the symmetry potential.

\begin{figure}[tbh]
\includegraphics[scale=0.75]{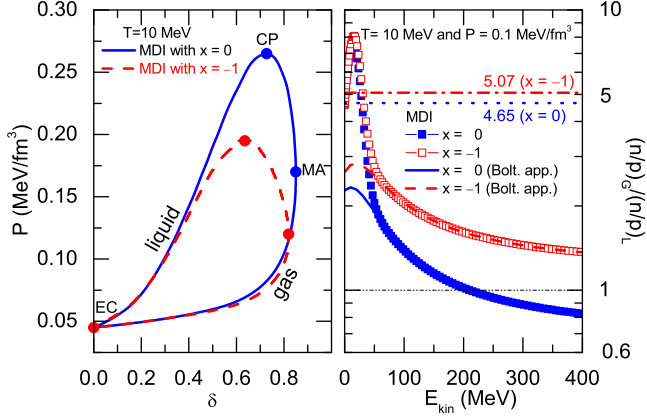}
\caption{(Color online) Left window: the section of binodal surface
at $T=10$ MeV from the MDI interaction with $x=0$ and $x=-1$. Right window:
the double neutron/proton ratio in the gas and liquid phases
$(n/p)_{G}/(n/p)_{L}$ as a function of the nucleon kinetic energy.
Taken from Ref.~\cite{LiBA07}.}
\label{DiffIsoFrac}
\end{figure}

As an example, a typical section of the binodal surface at $T=10 $ MeV
is shown in the left window of Fig.~\ref{DiffIsoFrac} using the MDI
interaction with $x=0$ and $x=-1$. The integrated IsoF phenomenon
with a more neutron-rich gas phase is clearly observed. Moreover,
one can see that the stiffer symmetry energy ($x=-1$) significantly
lowers the critical point (CP). When the pressure is less than than
about $P=0.12\text{ MeV}$/fm$^{3}$, the magnitude of the integrated
IsoF becomes, however, essentially independent of the stiffness of the
symmetry energy.

To demonstrate the advantages of the differential IsoF analyses over the
integrated ones, we study the differential IsoF at $T=10$ MeV and
$P=0.1\text{ MeV}$/fm$^{3}$ for which the integrated IsoF is essentially
independent of the $x$ parameter as seen in the left window of Fig.~\ref{DiffIsoFrac}.
Shown in the right window of Fig.~\ref{DiffIsoFrac} are the double
neutron/proton ratios in the gas and liquid phases $\frac{(n/p)_{G}}{(n/p)_{L}}(p)$
as a function of nucleon momentum or kinetic energy, i.e., the differential IsoFs,
for both $x=0$ and $x=-1$ in the MDI interaction. Interestingly, one can
see that the double neutron/proton ratios in both cases exhibit a strong momentum
dependence. In addition, while the integrated double neutron/proton
ratios of $4.65$ ($x=0$) and $5.07$ ($x=-1$) are very close to each
other, the differential IsoF for nucleons with kinetic energies higher
than about $50$ MeV is very sensitive to the $x$ values used for the
density dependence of the symmetry energy. Furthermore, it is surprising
to see that the IsoF for $x=0$ becomes less than one for nucleons with kinetic energies
higher than about $220$ MeV, and this means there are more energetic neutrons
than protons in the liquid phase compared to the gas phase.

Experimentally, it could be a big challenge to measure the differential
IsoF because the momentum distribution of the neutron/proton ratio
in the liquid phase may not be directly measured since only free nucleons
and bound nuclei in their ground states at the end of the collisions can
be detected in heavy ion collisions. Nevertheless, precursors and/or residues
of the effects due to the differential IsoF may still be detectable in
heavy ion collisions induced by radioactive beams~\cite{LiBA07}. While
it may be very challenging to test experimentally the predictions of the
differential IsoF, future comparisons with experimental data will allow
us to extract critical information on the nuclear symmetry potential,
especially its momentum dependence, and thus give deeper insight on the
isospin- and momentum-dependent effective interactions.

\subsection{The inner edge of neutron star crust}

Neutrons stars provide an excellent site to explore the properties of
nuclear matter at extreme isospin conditions and thus become an imprtant
astrophysical laboratory to investigate the isospin- and
momentum-dependent in-medium nuclear effective interactions. In the
present section, we highlight some results obtained with the MDI
interaction on the location of the inner edge of neutron star
crust. The latter separates the liquid core from the inner crust in
neutron stars, and it plays an important role in determining the
structural properties of neutron stars such as the crustal fraction
of total moment of inertia and the mass-radius relations of static
neutron stars~\cite{Lat00}.

The transition density $\rho _{t}$ at the inner edge of neutron star
crust can be determined from comparing relevant properties of the
nonuniform solid crust and the uniform liquid core. This is, however,
very difficult in practice since the inner crust may contain the
so-called \textquotedblleft nuclear pasta\textquotedblright\ with very
complicated geometries~\cite{Lat00,Rav83,Oya93,Hor04,Ste08}. A good
approximation used in the determination of $\rho _{t}$ is to search
for the density at which the uniform liquid first becomes unstable
against small amplitude density fluctuations with clusterization.
This approximation has been shown to give a very small error for the
actual core-crust transition density, and it would produce the exact
transition density for a second-order phase
transition~\cite{Pet95b,Dou00,Dou01,Hor03}. In
the literature, several such methods, including the thermodynamical
method~\cite{Kub07,Lat07,Wor08}, the dynamical curvature matrix
method~\cite{Pet95b,Dou00,BPS71,BBP71,Pet95a,Oya07,Duc07}, the
Vlasov equation method~\cite{chomaz,Pro06,Duc08a,Duc08b,Pai10}, and
the random phase approximation (RPA)~\cite{Hor01,Hor03,Duc08a}, have
been used extensively. Once the $\rho _{t}$ is determined, one can
easily obtain the pressure $P_t$ at the inner edge, which is also
an important quantity and might be measurable indirectly from
observations of pulsar glitches~\cite{Lat07,Lin99}. In the following,
we briefly introduce the dynamical (curvature matrix) method
and present some results obtained with this method.

In the dynamical method, a homogeneous $npe$ matter will be stable
against small periodic density perturbations with clusterization if
the following condition can be satisfied~\cite{Pet95b,BPS71,BBP71,Pet95a,Oya07}
\begin{equation}
V_{\rm dyn}(k)=V_{0}+\beta k^{2}+\frac{4\pi e^{2}}{k^{2}+k_{TF}^{2}}>0,
\label{Vdyn}
\end{equation}%
with
\begin{eqnarray}
V_{0} &=&\frac{\partial \mu _{p}}{\partial \rho _{p}}-\frac{(\partial \mu
_{n}/\partial \rho _{p})^{2}}{\partial \mu _{n}/\partial \rho _{n}},\text{ }%
k_{TF}^{2}=\frac{4\pi e^{2}}{\partial \mu _{e}/\partial \rho _{e}},  \notag \\
\beta  &=&D_{pp}+2D_{np}\zeta +D_{nn}\zeta ^{2},~~\zeta
=-\frac{\partial \mu _{n}/\partial \rho _{p}}{\partial \mu
_{n}/\partial \rho _{n}}.  \notag
\end{eqnarray}%
In the above expressions, $\mu _{i}$ is the chemical potential of
particle $i$ and $k$ is the wavevector of the spatially periodic
density perturbations. The three terms in Eq.~(\ref{Vdyn})
represent the contributions from the bulk nuclear matter, the
density gradient (surface) terms, and the Coulomb interaction,
respectively. $D_{np}=D_{pn}$ and $D_{pp}=D_{nn}$ are coefficients
of density gradient terms. The $V_{\rm dyn}(k)$ has the minimal
value of $V_{\rm dyn}(k_{\min })=V_{0}+2(4\pi e^{2}\beta )^{1/2}-\beta k_{TF}^{2}$
at $k_{\min }=[(\frac{4\pi e^{2}}{\beta })^{1/2}-k_{TF}^{2}]^{1/2}$~\cite%
{BPS71,BBP71,Pet95a,Pet95b,Oya07}. The density at which $V_{\rm dyn}(k_{\min })$
vanishes then corresponds to the $\rho _{t}$.

\begin{figure}[t!]
\includegraphics[scale=0.78]{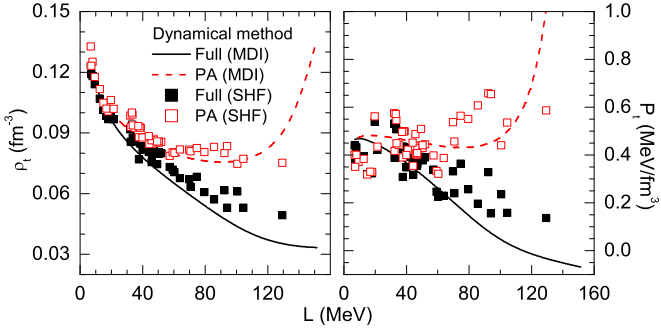}
\caption{(Color online) The transition density $\rho _{t}$ (left
panel) and the corresponding pressure $P_{t}$ (right panel) as
functions of symmetry energy slope parameter $L$ by using the
dynamical methods with the full EOS and its parabolic approximation
(PA). The MDI (curves) and Skyrme (squares) interactions are used.
The calculated results are taken from Refs.~\cite{Xu09a,Xu09b}.}
\label{RhotPtL}
\end{figure}

Shown in Fig.~\ref{RhotPtL} are the $\rho _{t}$ and the
corresponding $P_{t}$ as functions of the slope parameter $L$ of the
symmetry energy using the dynamical methods with the MDI interaction
by varying $x$ and $51$ Skyrme forces (See Ref.~\cite{Xu09b} for the
details of the $51$ Skyrme forces). For comparisons, we have
included results with the full EOS and its parabolic approximation
(PA), i.e., $E(\rho ,\delta )=E(\rho ,\delta =0)+E_{\rm sym}(\rho
)\delta ^{2}+O(\delta^{4})$ from the same MDI interaction and Skyrme
forces. For the MDI interaction, which does not have gradient terms
in its present form, the coefficients of density gradient terms
have been assumed to be $D_{pp}=D_{nn}=D_{np}=132$ MeV$\cdot $fm$^{5}$,
which are consistent with the Skyrme-Hartree-Fock calculations~\cite{Oya07,Xu09a}.
With the full EOS, it is clearly seen that the $\rho _{t}$ and $P_{t}$
decrease quickly with increasing $L$ for both MDI and Skyrme
interactions with the former giving lower values than the latter.
On the other hand, it is very surprising to see that the PA drastically
changes the results, especially for stiffer symmetry energies
(i.e., larger $L$ values). Actually, the large error introduced by the
PA can be understood from the fact that the $\beta $-stable $npe$ matter
is usually highly neutron-rich and the higher-order $\delta $ term
contribution is appreciable. This is especially true for the stiffer
symmetry energy (i.e., larger $L$ values) which generally gives rise
to a more neutron-rich $npe$ matter at subsaturation densities. In addition,
since the energy curvatures are involved in the stability conditions, the
contributions from higher-order terms in the EOS are thus multiplied by a
larger factor than the quadratic term~\cite{Xu09a,Xu09b}. These features
are in agreement with the early finding~\cite{Arp72} that the $\rho_{t}$
is very sensitive to the details of the nuclear EOS.

The above results indicate that to determine the $\rho_{t}$ and $P_{t}$,
one may introduce a big error by assuming {\it a priori} that the EOS of
asymmetric nuclear matter is parabolic in $\delta $ for a given interaction.
Therefore, the $\rho_t$-$L$ and $P_t$-$L$ correlation obtained with the
full EOS should be used to constrain the $\rho_{t}$ and $P_t$ from the
experimentally constrained $L$ values. From Fig.~\ref{RhotPtL}, one
can obtain $\rho_{t} \approx 0.075 \pm 0.015$ fm$^{-3}$ and $P_{t}
\approx 0.37 \pm 0.17$ MeV/fm$^3$ if one assumes a constraint of
$L=50 \pm 20$ MeV which probably represents our present best knowledge on
the $L$ parameter~\cite{ChenLW12}. The present results also demonstrate
that the isospin- and momentum-dependent effective interactions play
a critically important role in determining the core-crust transition density
$\rho _{t}$ and the corresponding pressure $P_{t}$ in neutron stars.

It should be mentioned that the isospin- and momentum-dependent effective
interactions may also significantly affect the critical density above which
the proton fraction is larger than about $1/9$ so as to trigger the direct
URCA process that can lead to a faster cooling of neutron stars~\cite{Lat91}.
From Fig. 2 of Ref.~\cite{Xu09b}, we find that the critical density of the
direct URCA process is about $0.25$ ($0.55$) fm$^{-3}$ for the MDI
interaction with $x=-1$ ($x=0$). Furthermore, the central density of a
neutron star with canonical mass of $1.4  M_{\odot}$ is found to be about
$0.42$ ($0.57$) fm$^{-3}$ for the MDI interaction with $x=-1$ ($x=0$).
These results imply that the direct URCA process may occur in a neutron
star with canonical mass of $1.4  M_{\odot}$ for the MDI interaction
with $x=-1$ and  $x=0$.

\section{Extension of the MDI interaction}
\label{SectionExtension}

While the MDI interaction has been extensively applied in transport model
simulations of heavy ion collisions as well as the investigations of thermal
properties of asymmetric nuclear matter and neutron stars, some studies
have also been performed to extend and improve the MDI interaction in recent
years. In this section, we review the extended MDI interaction~\cite{XuJ10a}
for the baryon octet and its application to hybrid stars as well as the
improved MDI interaction~\cite{XuC10a} with separate density-dependent
terms for neutrons and protons to take into account more accurately
the effect of the isospin dependence of in-medium many-body forces.

\subsection{The extended MDI interaction for the baryon octet}

During the past decades, significant progress has been made in
understanding the in-medium effective NN interaction. In contrast,
the nucleon-hyperon (NY) and hyperon-hyperon (YY) interactions in
nuclear medium are poorly known. The latter is important for
understanding a number of important issues in nuclear physics and
astrophysics, such as the properties of hypernuclei, the
production of strange hadrons in high energy heavy ion collisions,
the EOS of dense baryonic matter, and the properties of neutron
stars that may have abundant hyperons in their interiors. Therefore,
it is of great importance to develop an effective model for the NY
and YY interactions in nuclear medium. In the following, we review
the work on extending the MDI interaction to include NY and YY
interactions~\cite{XuJ10a}.

\subsubsection{The extended MDI interaction}

In the extended MDI interaction~\cite{XuJ10a}, the NY and YY
interactions have been assumed to have the same density and
momentum dependence as the interactions between two nucleons
in the MDI interaction. The potential energy density of a
hypernuclear matter due to interactions between any
two octet baryons then has the following form:
\begin{eqnarray}
V_{bb^\prime} &=& \sum_{\tau_b,\tau^\prime_{b^\prime}}\left[
\frac{A_{bb^\prime}}{2 \rho_0} \rho_{\tau_b}
\rho_{\tau^\prime_{b^\prime}} + \frac{A^\prime_{bb^\prime}}{2
\rho_0} \tau_b \tau_{b^\prime}^\prime \rho_{\tau_b}
\rho_{\tau^\prime_{b^\prime}}\right. \notag\\
&+& \left.\frac{B_{bb^\prime}}{\sigma+1}
\frac{\rho^{\sigma-1}}{\rho^\sigma_0}(\rho_{\tau_b}
\rho_{\tau^\prime_{b^\prime}} - x \tau_b \tau_{b^\prime}^\prime
\rho_{\tau_b} \rho_{\tau^\prime_{b^\prime}})\right. \notag\\
&+& \left.\frac{C_{{\tau_b},{\tau^\prime_{b^\prime}}}}{\rho_0} \int
\int d^{3}{\vec p}d^{3}{\vec p}^{\prime }\frac{f_{\tau_b}(\vec{r},\vec{p}%
)f_{\tau^\prime_{b^\prime}}(\vec{r},\vec{p}^{\prime
})}{1+(\vec{p}-\vec{p}^{\prime })^{2}/\Lambda ^{2}}\right],
\label{Vbb}
\end{eqnarray}
where $b$ ($b^\prime$) denotes the baryon octet, i.e., $N$,
$\Lambda$, $\Sigma$, and $\Xi$. We use the conventions that
$\tau_N=-1$ for neutron and $1$ for proton (Note: this is opposite
to the convention used earlier), $\tau_\Lambda=0$ for $\Lambda$,
$\tau_\Sigma=-1$ for $\Sigma^-$, $0$ for $\Sigma^0$ and $1$ for
$\Sigma^+$, and $\tau_\Xi=-1$ for $\Xi^-$ and $1$ for $\Xi^0$. In
the above, the total baryon density is then given by
$\rho = \sum_b \sum_{\tau_b} \rho_{\tau_b}$, and
$f_{\tau_b}({\vec r},{\vec p})$ is the phase-space distribution
function of particle species $\tau_b$. The $A_{bb^\prime}$,
$A^\prime_{bb^\prime}$, $B_{bb^\prime}$, and $C_{\tau_b,\tau_b^\prime}$
are interaction parameters. If there are only nucleons, one can
rewrite $A_{NN}=(A_l+A_u)/2$, $A^\prime_{NN}=(A_l-A_u)/2$,
$B_{NN}=B$, and $C_{{\tau_N},{\tau^\prime_{N}}}=C_l$ for
$\tau_N=\tau^\prime_{N}$ and $C_{{\tau_N},{\tau^\prime_{N}}}=C_u$
for $\tau_N \ne \tau^\prime_{N}$, which then reduce to the original
parameters in the MDI interaction for nucleons~\cite{Das03,Che05a}.
Again, the parameter $x$ is used here to model the isospin effect
on the interaction energy in hypernuclear matter.

The single-particle potential for a baryon of species $\tau_b$ in a
hypernuclear matter can then be obtained from the total potential
energy density of the hypernuclear matter, given by $V_{HP}
=(1/2)\sum_{b,b^\prime} V_{bb^\prime}$, as
\begin{eqnarray}
U_{\tau_b}(p) &=& \frac{\delta}{\delta \rho_{\tau_b}}
V_{HP} \notag\\
&=&\sum_{b^\prime(b^\prime \ne
b)}\sum_{\tau^\prime_{b^\prime}}\left [ \frac{A_{bb^\prime}}{2
\rho_0} \rho_{\tau^\prime_{b^\prime}} +
\frac{A^\prime_{bb^\prime}}{2 \rho_0} \tau_b \tau^\prime_{b^\prime}
\rho_{\tau^\prime_{b^\prime}} \right.\notag\\
&+& \left.\frac{B_{bb^\prime}}{\sigma+1}
\frac{\rho^{\sigma-1}}{\rho^\sigma_0} (\rho_{\tau^\prime_{b^\prime}}
- x \tau_b \tau^\prime_{b^\prime} \rho_{\tau^\prime_{b^\prime}}) +
\frac{C_{\tau_b,\tau^\prime_{b^\prime}}}{\rho_0}
\right.\notag\\
&\times& \left. \int d^{3}{\vec p}^{\prime }
\frac{f_{\tau^\prime_{b^\prime}}(\vec{r},\vec{p}^{\prime
})}{1+(\vec{p}-\vec{p}^{\prime })^{2}/\Lambda ^{2}} \right]
+\sum_{\tau^\prime_{b}} \left[ \frac{A_{bb}}{\rho_0}
\rho_{\tau^\prime_{b}} \right.\notag\\
&+& \left.\frac{A^\prime_{bb}} {\rho_0} \tau_b \tau^\prime_{b}
\rho_{\tau^\prime_{b}} + \frac{2 B_{bb}}{\sigma+1}
\frac{\rho^{\sigma-1}}{\rho^\sigma_0}( \rho_{\tau^\prime_{b}} - x
\tau_b \tau^\prime_{b} \rho_{\tau^\prime_{b}}) \right.\notag\\
&+& \left.\frac{2 C_{\tau_b,\tau^\prime_{b}}}{\rho_0} \int
d^{3}{\vec p}^{\prime } \frac{f_{\tau^\prime_{b}}(\vec{r},\vec{p}^{\prime
})}{1+(\vec{p}-\vec{p}^{\prime })^{2}/\Lambda ^{2}} \right] \notag\\
&+& \sum_{b^\prime,b^\prime{^\prime} } \left[ B_{b^\prime
b^\prime{^\prime}}
\frac{\sigma-1}{\sigma+1}\frac{\rho^{\sigma-2}}{\rho^\sigma_0}
\right.\notag\\
&\times& \left.\sum_{\tau_{b^\prime}}
\sum_{\tau^\prime_{b^\prime{^\prime}}} (\rho_{\tau_{b^\prime}}
\rho_{\tau^\prime_{b^\prime{^\prime}}} - x \tau_{b^\prime}
\tau^\prime_{b^\prime{^\prime}} \rho_{\tau_{b^\prime}}
\rho_{\tau^\prime_{b^\prime{^\prime}}}) \right]. \label{Ubb}
\end{eqnarray}

The parameters $A_{bb^\prime}$, $A^{\prime}_{bb^\prime}$, $B_{bb^\prime}$,
and $C_{{\tau_b},{\tau^\prime_{b^\prime}}}$ for NY and YY interactions
can in principle be determined from the free space NY and YY interactions.
Due to the lack of NY scattering experiments, knowledge on the
NY interactions has been mainly obtained from the hyperon single-particle
potentials extracted empirically from analyzing $\Lambda$~\cite{Has06} as
well as $\Sigma$~\cite{Dover89,Bart99} and $\Xi$~\cite{Dover83} production
in nuclear reactions. Although the NY interaction has been extensively
studied in the past~\cite{Maessen89,Reuber94,Dabrowski99}, the isospin and
momentum dependence of the in-medium NY interactions are still not well
determined, and the situation is even worse for YY interactions. In the
extended MDI interaction, the parameters $A_{bb^\prime}$, $A^{\prime}_{bb^\prime}$,
$B_{bb^\prime}$, and $C_{{\tau_b},{\tau^\prime_{b^\prime}}}$
are thus assumed to be proportional to corresponding ones in the NN interaction.
Particularly, for $A_{bb^\prime}$, $A^{\prime}_{bb^\prime}$, and $B_{bb^\prime}$,
one has
\begin{eqnarray}
A_{bb^\prime} &=& f_{bb^\prime} A_{NN}, \notag\\
A^\prime_{bb^\prime} &=& f_{bb^\prime} A^\prime_{NN}, \notag\\
B_{bb^\prime} &=& f_{bb^\prime} B_{NN},
\end{eqnarray}
and for $C_{{\tau_b},{\tau^\prime_{b^\prime}}}$, one has
\begin{displaymath}
C_{{\tau_b},{\tau^\prime_{b^\prime}}} = \left\{\begin{array}{ll}
f_{bb^\prime} \frac{C_l+C_u}{2} &({\tau_b} ~\text{or}~ {\tau^\prime_{b^\prime}} = 0),\\
f_{bb^\prime} C_l &({\tau_b} = {\tau^\prime_{b^\prime}} \ne 0),\\
f_{bb^\prime} C_u &({\tau_b} \ne {\tau^\prime_{b^\prime}} \ne
0),\end{array} \right.
\end{displaymath}
with hyperons $\Lambda$ and $\Sigma^0$ treated differently.

\begin{table*}[tbp]
\caption{Parameters for the MDI-Hyp-A and MDI-Hyp-R interactions
with $x=0$ and $x=-1$. All except $\sigma$ are in units of MeV.
$A_{N\Sigma}^\prime(R)$ and $B_{N\Sigma}(R)$ are for the MDI-Hyp-R
interaction, and $A_{N\Sigma}^\prime(A)$ and $B_{N\Sigma}(A)$ are
for the MDI-Hyp-A interaction. Other parameters are the same for
both interactions. Taken from Ref.~\cite{XuJ10a}.}
\label{para}
\centering
\begin{tabular}{ccccccccccccc}
\hline\hline $A_{NN}$ & $A_{N\Lambda}$ & $A_{N\Sigma}$ & $A_{N\Xi}$
& $A_{\Lambda\Lambda}$ & $A_{\Lambda\Sigma}$ & $A_{\Lambda\Xi}$ &
$A_{\Sigma\Sigma}$ &
$A_{\Sigma\Xi}$ & $A_{\Xi\Xi}$ & $\Lambda$ & $\sigma$\\
$$ -108.28 & -108.28 & -108.28 & -79.04 & -68.21 & -135.34 & -135.34 & -53.05 & -108.28 & -57.39 & 263.04 & 4/3\\
\hline
$x$ & $A_{NN}^\prime$ & $A_{N\Sigma}^\prime(A)$ & $A_{N\Sigma}^\prime(R)$ & $A_{N\Xi}^\prime$ & $A_{\Sigma\Sigma}^\prime$ & $A_{\Sigma\Xi}^\prime$ & $A_{\Xi\Xi}^\prime$ \\
$$ 0 & -12.29 & -12.29 & -28.65 & -8.98 & -6.02 & -12.29 & -6.52\\
$$ -1 & -103.45 & -103.45 & -241.04 & -75.52 & -50.69 & -103.45 & -54.83 \\
\hline
 $B_{NN}$ & $B_{N\Lambda}$ & $B_{N\Sigma}(A)$ &
$B_{N\Sigma}(R)$ & $B_{N\Xi}$ &  $B_{\Lambda\Lambda}$ & $B_{\Lambda\Sigma}$ & $B_{\Lambda\Xi}$ & $B_{\Sigma\Sigma}$ & $B_{\Sigma\Xi}$ & $B_{\Xi\Xi}$\\
$$ 106.35 & 106.35 & 106.35 & 247.80 & 77.64 & 67.00 & 132.94 & 132.94 & 52.11 & 106.35 & 56.37 \\
\hline
 $C_{\tau_N,\tau_N}$ & $C_{\tau_N,-\tau_N}$ &
$C_{\tau_N,\tau_\Sigma}$ & $C_{\tau_N,-\tau_\Sigma}$ &
$C_{\tau_N,\tau_\Xi}$ & $C_{\tau_N,-\tau_\Xi}$ &
$C_{\tau_\Sigma,\tau_\Sigma}$ & $C_{\tau_\Sigma,-\tau_\Sigma}$ &
$C_{\tau_\Sigma,\tau_\Xi}$ & $C_{\tau_\Sigma,-\tau_\Xi}$ &
$C_{\tau_\Xi,\tau_\Xi}$ &
 $C_{\tau_\Xi,-\tau_\Xi}$ \\
$$ -11.70 & -103.40  & -11.70 & -103.40 & -8.54 &
-75.48 & -5.73 & -50.67 & -11.70 & -103.40 & -6.20 & -54.80\\
\hline $C_{N\Lambda}$ & $C_{N\Sigma^0}$ &  $C_{\Lambda\Lambda}$ &
$C_{\Lambda\Sigma}$ & $C_{\Lambda\Xi}$ & $C_{\Sigma^0\Sigma}$ &
 $C_{\Sigma^0\Xi}$\\
$$  -57.55 & -57.55 & -36.26 & -71.94 & -71.94 & -28.20 & -57.55\\
\hline\hline
\end{tabular}%
\end{table*}

The values of $f_{bb^\prime}$ are determined by fitting the
empirical potential $U^{(b^\prime)}_b$ of baryon $b$ at rest in a
medium consisting of baryon species $b^\prime$. For hyperons in
symmetric nuclear matter at saturation density, their potentials are
\begin{equation}\label{fitlambda}
U^{(N)}_\Lambda(\rho_N^{} = \rho_0^{}) = - 30 ~\text{MeV}
\end{equation}
for the $\Lambda$ potential from the analysis of $(\pi^+, K^+)$
and $(K^-, \pi^-)$ reactions~\cite{Millener88,Chrien89} and
\begin{equation}\label{fitXi}
U^{(N)}_\Xi(\rho_N^{} = \rho_0^{}) = - 18 ~\text{MeV}
\end{equation}
for the $\Xi$ potential from the analysis of
$(\Xi,~^4_{\Lambda}H)$~\cite{Aok95} and $(K^-,
K^+)$~\cite{Fukuda98,Khaustov00} reactions. This leads to
$f_{N\Lambda} = 1$ and $f_{N\Xi} = 0.73$. For the $\Sigma$ hyperon,
its potential in symmetric nuclear matter at saturation density was
taken to be attractive in earlier studies~\cite{Dover89}, but more
recent analysis indicate that it should be
repulsive~\cite{Batty94,Mares95,Noumi02,Har05,Friedman07}. To take
into account these uncertainties, both the attractive and repulsive
cases
\begin{eqnarray}\label{fitsigma}
U^{(N)}_\Sigma(\rho_N^{} = \rho_0^{}) &=& \pm 30 ~\text{MeV}
\end{eqnarray}
have therefore been considered in Ref.~\cite{XuJ10a}. By setting
$f_{N\Sigma}=1$ one obtains an attractive $\Sigma$N interaction,
called MDI-Hyp-A in the following. To get a repulsive $\Sigma$N
interaction, called MDI-Hyp-R in the following, one can adjust
the values of positive and negative terms in the single-particle
potential by setting $B_{N\Sigma} = 2.33 B_{NN}$ and
$A_{N\Sigma}^\prime = 2.33A_{NN}^\prime$ but keeping other
parameters as in the MDI-Hyp-A interaction. A similar method of changing
an attractive $\Sigma$N interaction to a repulsive one was used in
the relativistic mean-field model calculation~\cite{Schaffer96}
by adjusting the coupling constants of $\omega$ and $\rho$ mesons.

For the YY interaction, the parameters are determined according
to~\cite{Schaffer94}
\begin{equation}
U^{(Y^\prime)}_Y(\rho_{Y^\prime}^{} = \rho_0^{}) \sim
-40~\text{MeV},
\end{equation}
which leads to
$f_{\Lambda\Lambda}=0.63$, $f_{\Lambda\Sigma}=1.25$,
$f_{\Lambda\Xi}=1.25$, $f_{\Sigma\Sigma}=0.49$, $f_{\Sigma\Xi}=1$,
and $f_{\Xi\Xi}=0.53$ for the strength of the YY interactions.

For completeness, we list the detailed parameter values of
the MDI-Hyp-A and MDI-Hyp-R interactions with $x=0$ and $x=-1$ in
Table.~\ref{para}. These parameterizations can be considered as a
baseline for studying the properties of hypernuclear matter, and
more sophisticated treatments can be made in future after the
in-medium properties of hyperons are better understood.

\begin{figure}[ht]
\centerline{\includegraphics[scale=0.9]{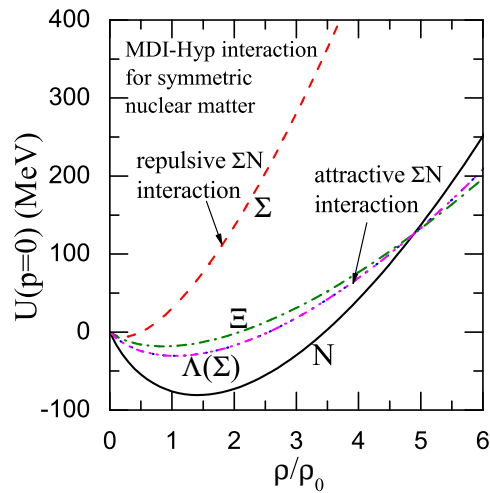}}
\caption{(Color online) Density dependence of single-particle
potentials for particles at rest in symmetric nuclear matter.
Taken from Ref.~\cite{XuJ10a}.}
\label{SPPrhoH}
\end{figure}

\subsubsection{Single-particle potentials for the baryon octet in asymmetric nuclear matter}

The single-particle potential as given by Eq.~(\ref{Ubb}) is an important
quantity linked to the interaction of a particle in nuclear medium. Shown
in Fig.~\ref{SPPrhoH} is the density dependence of the single-particle
potential of a particle at rest in symmetric nuclear matter. One can see
that, although the nucleon potential is more attractive at saturation
density $\rho_0 $ than those of hyperons, it becomes more repulsive than
the hyperon potentials above about $5\rho_0 $, including the $\Sigma$
potential with MDI-Hyp-A. For the $\Sigma$ potential with MDI-Hyp-R, it
becomes more repulsive with increasing density and becomes weakly
attractive only at very low densities. Compared with results from other
models given in Ref.~\cite{Dap09} (and references therein), the
single-particle potentials of $\Lambda$ and $\Sigma$ presented here are
close to those from the chiral EFT~\cite{Pol06}, but more repulsive than
those from the G-matrix calculations within the soft core Nijmegen model
or the J$\rm{\ddot{u}}$lich meson-exchange model for the free NY
interactions, specially at high densities.

\begin{figure}[ht]
\centerline{\includegraphics[scale=0.9]{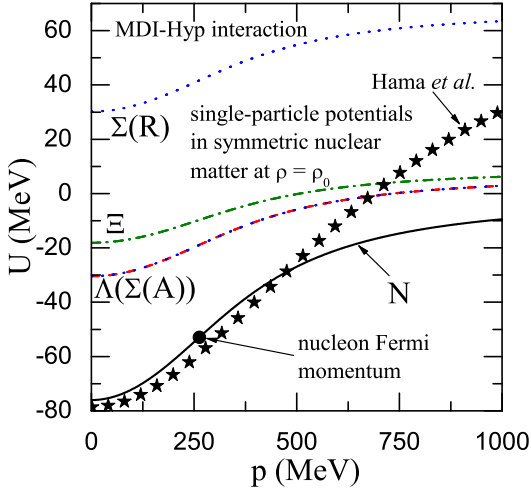}}
\caption{(Color online) Single-particle potentials in symmetric nuclear matter at
saturation density $\rho_0$ as functions of particle momentum. The
Schr$\ddot{o}$dinger equivalent potential obtained by Hama {\it et
al.}~\cite{Hama90,Coo93} from the nucleon-nucleus scattering data is
shown by stars for comparison. $\Sigma$(A) and $\Sigma$(R) are for
the MDI-Hyp-A and MDI-Hyp-R interactions, respectively. Taken from Ref.~\cite{XuJ10a}.}
\label{SPPrho0H}
\end{figure}

In the extended MDI interaction, the single-particle potential of
a particle is also momentum dependent. Shown in Fig.~\ref{SPPrho0H}
is the single-particle potential as a function of the particle momentum
for both nucleons and hyperons in symmetric nuclear matter at saturation
density $\rho_0 $. Again, results for $\Sigma $ potential with both
MDI-Hyp-A and MDI-Hyp-R interactions are shown for comparison. Also
indicated in Fig.~\ref{SPPrho0H} is the nucleon Fermi momentum. One can
see that the nucleon single-particle potential from the MDI
interaction is consistent with the Schr$\ddot{o}$dinger equivalent
potential obtained by Hama {\it et al.} from Dirac phenomenology
of the nucleon-nucleus scattering data~\cite{Hama90,Coo93} up to the
nucleon momentum of $500$ MeV. For hyperons, their single-particle
potentials from the extended MDI interaction agree with that obtained
from the G-matrix calculations with the free Nijmegen $NY$
interaction~\cite{Baldo98} at low momenta since both are constrained
by available experimental data. However, they are slightly different
at high momenta. Therefore, the momentum dependence of $NY$ and $YY$
interactions remains an open question, especially at high momenta.
In addition, the density dependence of the single-particle potential
for high momentum particles is poorly known.

\begin{figure}[ht]
\centerline{\includegraphics[scale=0.9]{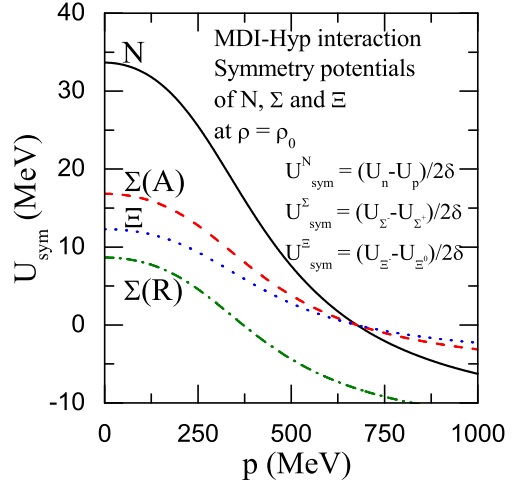}}
\caption{(Color online) Momentum dependence of the symmetry potentials
of nucleon, $\Sigma$ and $\Xi$ in asymmetric nuclear matter at saturation
density $\rho=\rho_0$ with the MDI-Hyp-A and MDI-Hyp-R interactions.
Taken from Ref.~\cite{XuJ10a}.}
\label{Usymrho0H}
\end{figure}

For the extended MDI interaction, similarly to the case of nucleons,
the single-particle potentials of $\Sigma$ and $\Xi$ in asymmetric
nuclear matter are also approximately linear in the isospin asymmetry
$\delta$ of the matter. The single-particle potential of a
particle in asymmetric nuclear matter can thus be approximated as
\begin{equation}
U_{\tau_b}(\rho, p, \delta)\approx
U_{\tau_b}(\rho, p, \delta=0)-\tau_b U^{b}_{\rm sym}(\rho, p) \delta,
\end{equation}
where the symmetry potential $U^b_{\rm sym}(\rho, p)$ can be
obtained approximately by
\begin{eqnarray}
U^N_{\rm sym}(\rho, p) &\approx &(U_n(\rho,p,\delta)-U_p(\rho,p,\delta))/2\delta,\\
U^{\Sigma}_{\rm sym}(\rho,p) &\approx & (U_{\Sigma^-}(\rho,p,\delta)-U_{\Sigma^+}(\rho,p,\delta))/2\delta,\\
U^{\Xi}_{\rm sym}(\rho,p) &\approx & (U_{\Xi^-}(\rho,p,\delta)-U_{\Xi^0}(\rho,p,\delta))/2\delta,
\end{eqnarray}
for the nucleon as well as $\Sigma$ and $\Xi$ hyperons, respectively.
Fig.~\ref{Usymrho0H} shows the momentum dependence of the symmetry
potentials for nucleon, $\Sigma$ and $\Xi$ at saturation density
$\rho_0 $ with the MDI-Hyp-A and MDI-Hyp-R interactions, and these
results are obtained from the single-particle potentials of nucleons
as well as $\Sigma$ and $\Xi$ hyperons in asymmetric nuclear matter
at $\rho_0 $ and isospin asymmetry $\delta=0.2$. One can see that
all symmetry potentials at $\rho_0 $ display strong momentum dependence
and decrease with increasing momentum.

\begin{figure}[h]
\centerline{\includegraphics[scale=0.9]{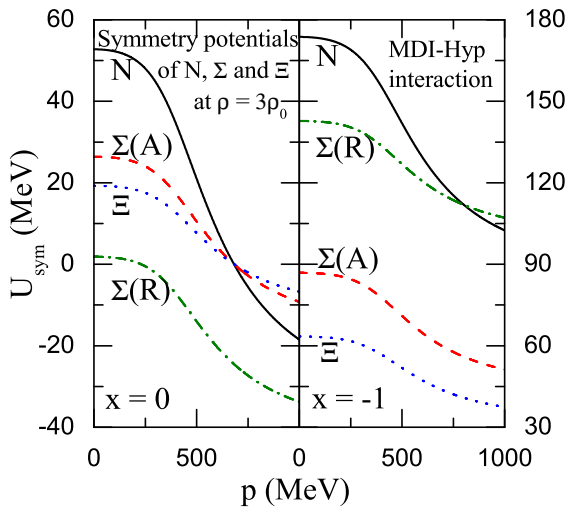}}
\caption{(Color online) Same as Fig.~\ref{Usymrho0H} but at a density
of $\rho=3\rho_0$ and with $x=0$ (left panel) and $x=-1$ (right panel).
Taken from Ref.~\cite{XuJ10a}.}
\label{Usym3rho0H}
\end{figure}

Although the symmetry potentials at saturation density $\rho_0$
are independent of the $x$ parameter by construction in the MDI
interaction as mentioned earlier, this is not the case at
other densities. Shown in Fig.~\ref{Usym3rho0H} are the momentum
dependence of the symmetry potentials of the nucleon as well as the
$\Sigma$ and $\Xi$ hyperons in asymmetric nuclear matter at density
$\rho=3\rho_0$ in the MDI-Hyp-A and MDI-Hyp-R interactions with
$x=0$ and $x=-1$. It is seen that the symmetry potentials depends
strongly on the particle momentum and the $x$ value used. For the
$\Sigma$ hyperon, its symmetry potential further depends on the choice
of the MDI-Hyp-A interaction or the MDI-Hyp-R interaction. It should
be mentioned that the charged $\Sigma$ baryon ratio in heavy ion
collisions has been proposed to constrain the symmetry energy
(potential) at densities larger than $3\rho_0$~\cite{QFLi05}.
Therefore, it will be very interesting to see how the symmetry
potentials of $\Sigma$ and $\Xi$ hyperons affect the charged
$\Sigma$ hyperon ratio and the charged $\Xi$ hyperon ratio in
intermediate and high energy heavy ion collisions induced by
neutron-rich nuclei. This may be important for accurately
constraining the high-density behavior of the symmetry energy
using these ratios in heavy ion collisions. Therefore, the
extended MDI interaction with hyperons is useful for
studying the nuclear symmetry energy (potential) at high densities
as well as the in-medium NN, NY, and YY effective interactions at
extreme conditions of high baryon densities, high momentum, and
high isospin, in transport model simulations for heavy ion collisions.

\subsubsection{Hybrid stars}

Besides heavy ion collisions in terrestrial laboratory, the compact
objects (e.g., neutron stars) provide another important site in nature
to test the in-medium effective interactions at high densities. Hyperons
may appear in the interior of neutron stars. At higher densities in the
core of a neutron star, a transition from the hadron matter to the quark
matter is also expected to occur. This leads to the so-called hybrid star
that is expected to have a hadron phase at low densities, a mixed phase
of hadrons and quarks at moderate densities, and a quark core at high
densities. The extended MDI interaction is thus useful for exploring the
properties of the hybrid star. In the following, we review the results of
the properties of hybrid stars with the extended MDI interaction.

\begin{figure}[h]
\centerline{\includegraphics[scale=0.9]{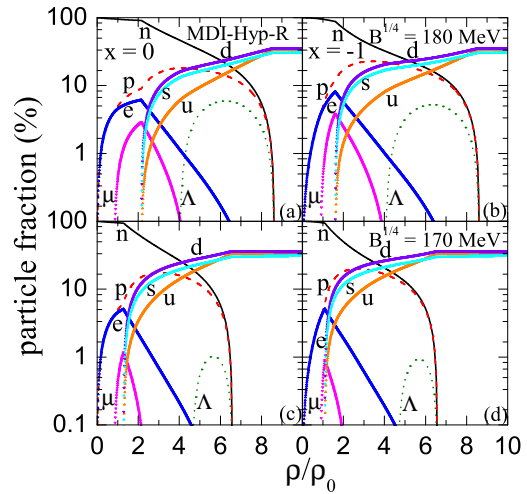}}
\caption{(Color online) Particle fractions as functions of baryon
density in a hypernuclear matter with the presence of a hadron-quark
phase transition from the MDI-Hyp-R interaction with $x=0$ ((a) and (c))
and $x=-1$ ((b) and (d)) for the hadron phase and the MIT bag model
for the quark phase. Results from $B^{1/4}=180$ MeV ((a) and (b)) and
$170$ MeV ((c) and (d)) are shown for comparison. Taken from Ref.~\cite{XuJ10a}.}
\label{MDIHQPTPF}
\end{figure}

For the hadron-quark phase transition, the Gibbs
construction~\cite{Glen92,Glen01} is adopted with the quark phase
described by a simple MIT bag model~\cite{Chodos74,Heinz86}. For the
hadron interactions, only the MDI-Hyp-R interaction is used in the
following since the repulsive $\Sigma$N interaction is more consistent
with the latest empirical
information~\cite{Batty94,Mares95,Noumi02,Har05,Friedman07}. Shown
in Fig.~\ref{MDIHQPTPF} is baryon density dependence of the particle
fractions of each species in the presence of a hadron-quark phase
transition, with the hadron phase described by the MDI-Hyp-R interaction
with $x=0$ and $x=-1$ and the quark phase described by the MIT bag model
with bag constants $B^{1/4}=180$ MeV and $170$ MeV. It is seen that
the hadron-quark phase transition occurs at lower baryon density for
a stiffer symmetry energy and for a smaller $B$ value, while the density
at the end of the hadron-quark phase transition essentially depend
only on the $B$ value but not much on the value of the $x$ parameter.
With a smaller $B$ value, the hadron-quark phase transition both begins
and ends at lower densities. In addition, one can see that only
$\Lambda$ hyperons (no other hyperons) appear in (and only in) the mixed
phase in the present model. However, the fraction of $\Lambda$ hyperons
is sensitive to the $B$ value with a smaller $B$ value giving smaller
fraction. Moreover, it should be mentioned that the fraction of
hyperons is also sensitive to the NY and YY interactions~\cite{XuJ10a}.
In particular, as shown in Fig. 6 of Ref.~\cite{XuJ10a}, the $\Lambda $
hyperon appears in hypernuclear matter at a baryon density of about
$0.5$ fm$^{-3}$ with the extended MDI interaction. For the $\Sigma $
hyperon, the critical density at which it appears in the hypernuclear
matter depends strongly on the sign of the $\Sigma N$ interaction.
For the attractive MDI-Hyp-A interaction, the critical density for the
appearance of $\Sigma^- $ is about $0.3$ fm$^{-3}$, whereas for the
repulsive MDI-Hyp-R interaction, it does not appear until very high
densities. These values are in good agreement with the values of about
$0.6$ for the $\Lambda$ hyperon and $0.3$ fm$^{-3}$ for the $\Sigma^-$
hyperon obtained from both the BHF+TBF~\cite{Bal00} and the DBHF~\cite{Sam10}
with free non-interacting hyperons.

For a static hybrid star, it contains three parts from the center to
the surface: the liquid core, the inner crust, and the outer crust.
The liquid core is assumed to be the hypernuclear matter or that with
the hadron-quark phase transition. For the inner crust, a parameterized
EOS of $P=a+b \epsilon^{4/3}$ is adopted as in the previous
treatment~\cite{Xu09a,Xu09b}. The well-known BPS EOS~\cite{BPS71} is
used for the outer crust which consists of heavy nuclei and the electron
gas. The transition density $\rho_{\rm t}$ between the liquid core and
the inner crust is consistently determined as described earlier (see also
Refs.~\cite{Xu09a,Xu09b}), and for the density at the edge of inner crust
and outer crust, it is taken to be
$\rho_{\rm out}=2.46\times10^{-4}$ fm$^{-3}$. The parameters $a$
and $b$ are determined by the pressures ($P$) and energy densities
($\epsilon$) at $\rho_{\rm t}$ and $\rho_{\rm out}$. Using these EOS's,
one can then calculate the mass-radius relation of hybrid stars with the
well-known Tolman-Oppenheimer-Volkoff (TOV) equation.

\begin{figure}[h]
\centerline{\includegraphics[scale=0.9]{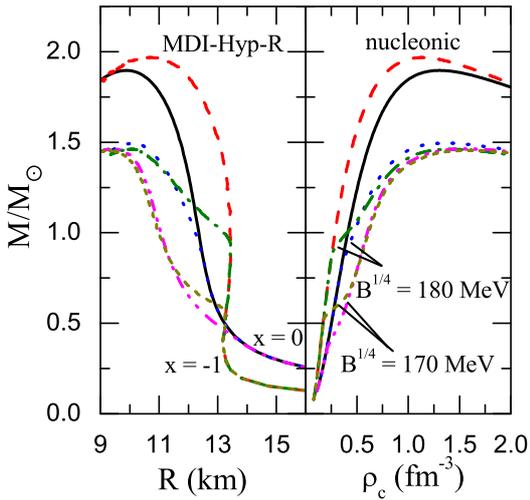}}
\caption{(Color online) The hybrid star mass as a function of radius (left panel)
and central density (right panel) using the MDI-Hyp-R interaction $x=0$ and $x=-1$
for the hadron phase and MIT bag constant $B^{1/4}=180$ and $170$ MeV for the
quark phase. Results from a pure nucleonic approach are also included for comparison.
Taken from Ref.~\cite{XuJ10a}.}
\label{MRrhocHQPT}
\end{figure}

Displayed in Fig.~\ref{MRrhocHQPT} are the M-R and M-$\rho_c$ relations
for hybrid stars using the MDI-Hyp-R interaction with $x=0$
and $x=-1$ for the hadron phase and $B^{1/4}=180$ and $170$ MeV for the
quark phase. For comparison, the results from a pure nucleonic approach
are also included. It is clearly seen that including the hadron-quark
phase transition in neutron stars significantly reduces the maximum mass
of neutron stars. For $B^{1/4}=180$ MeV, the maximum mass is
$1.50 M_{\odot}$ for $x=0$ and $1.46 M_{\odot}$ for $x=-1$, while for
$B^{1/4}=170$ MeV it is $1.46 M_{\odot}$ for $x=0$ and $1.45 M_{\odot}$
for $x=-1$, respectively. The radius of a standard neutron star with a
mass of $1.4 M_{\odot}$ is $11.0$ km for $x=0$ and $10.8$ km for $x=-1$
if $B^{1/4}=180$ MeV is used, while it becomes $10.2$ km for $x=0$ and
$10.0$ km for $x=-1$ if $B^{1/4}=170$ MeV is used. If the $B$ value is
further reduced, the hadron-quark phase transition would happen at even
lower densities, resulting in a smaller radius for the hybrid star.
These features indicate that the radius of a neutron star with canonical
mass $1.4 M_{\odot}$ is not only sensitive to the stiffness of the
symmetry energy, but to the hadron-quark phase transition.

Finally, we would like to point out that the original MDI interaction
does not cause causality violation in $\beta$-stable $npe\mu$
matter at least up to baryon density of $10\rho_0$ as shown in Fig. 2(d)
of Ref.~\cite{Xu09b}. Since the appearance of new degrees of freedom
such as hyperons and quarks usually softens the EOS of neutron star
matter, the causality condition is still satisfied for the extended
MDI interaction, and this is verified by explicit calculations~\cite{XuJ10a}. We also note that the causality condition
is satisfied for symmetric nuclear matter at least up to $10\rho_0$ for the MDI interaction.

\subsection{An improved MDI interaction with separate density
dependence for neutrons and protons}

In the MDI interaction, as pointed out before, the terms with
parameter B (and $\sigma$) in Eqs. (\ref{MDIV}) and (\ref{MDIU}) can
be obtained directly from the density-dependent two-body effective
interaction Eq. (\ref{MDI3BF}), which represents an effective
in-medium many-body force and has been extensively adopted in the
Skyrme or Gogny interaction. Since the pp, nn, and np interactions
in Eq. (\ref{MDI3BF}) all depend on the same total density
$\rho=\rho_n+\rho_p$, the proper isospin dependence of the
in-medium effective many-body forces is thus neglected. As pointed
out earlier in Refs.~\cite{koh,Dut1,Dut2,Dut3,Dut4,Dut5,Dut6},
there is no {\it a priori} physical justification for such a density
dependence in these interactions. On the other hand, within the
Brueckner theory, it has been found that the G-matrix of NN
interactions in isospin asymmetric nuclear matter depends strongly
on the respective Fermi momenta of neutrons and protons ($k_n$ and
$k_p$)~\cite{bru64,Dab73}. It is thus physically more reasonable to
assume that the interaction between neutrons depends on the neutron
density, and that between protons on the proton density, instead of
the total density $\rho$~\cite{koh}.

Indeed, the separate density dependence for pp, nn and np interactions
has already been used in various models to better understand the
structure of nuclei far from the $\beta$-stable line. For example,
in the early 1960s and 1970s, local effective interactions with density
dependence separately introduced for pp, nn, and np pairs were proposed
by Sprung and Banerjee~\cite{Spr}, Brueckner and Dabrowski~\cite{bru64,Dab73},
and Negele~\cite{Neg}. Recently, Xu and Li~\cite{XuC10a} explored the
effects of separate density dependence of neutrons and protons by
replacing the density-dependent term in Eq.~(\ref{MDI3BF}) with the
following expression:
\begin{eqnarray}\label{VD}
V_D=\frac{1}{6}t_3 (1+x_3 P_{\sigma}) [\rho_{\tau_i}(\textbf{r}_i) +
\rho_{\tau_j}(\textbf{r}_j)]^{\gamma} \delta(\textbf{r}_{ij}),
\label{MDI3BFnp}
\end{eqnarray}
where $\rho_{\tau}(\textbf{r})$ denotes the density of nucleon
$\tau$ ($1$ for neutrons and $-1$ for protons) at the coordinate
$\textbf{r}$. This density-dependent interaction changes
the expressions for the terms with parameter $B$ in the original MDI
interaction. In particular, the $B$-term in Eq.~(\ref{MDIV}) for the
energy density, i.e.,
\begin{eqnarray}
V_B &=&\frac{B}{\sigma +1}\frac{\rho ^{\sigma +1}}{\rho _{0}^{\sigma
}}(1-x\delta ^{2}) \label{MDIVB}
\end{eqnarray}%
becomes
\begin{eqnarray}
V'_B &=&\frac{B}{\sigma +1}\frac{\rho
^{\sigma +1}}{\rho _{0}^{\sigma }}\bigg\{\frac{1+x}{2}(1-\delta ^{2}) \notag \\
&+& \frac{1-x}{4}[(1+\delta)^{\sigma +1}+(1-\delta)^{\sigma
+1}]\bigg\}, \label{MDIVBpr}
\end{eqnarray}%
and the $B$-term in Eq. (\ref{MDIU}) for the single-particle
potential, i.e.,
\begin{eqnarray}
U_B &=&B\left(\frac{\rho }{\rho _{0}}\right)^{\sigma }(1-x\delta ^{2})-4\tau x\frac{B}{%
\sigma +1}\frac{\rho ^{\sigma -1}}{\rho _{0}^{\sigma }}\delta \rho
_{-\tau } \label{MDIB}
\end{eqnarray}%
becomes
\begin{eqnarray}
U'_B &=&\frac{B}{2}\bigg(\frac{2\rho_\tau }{\rho _{0}}\bigg)^{\sigma }(1-x) + \frac{2B}{%
\sigma +1}\bigg(\frac{\rho }{\rho _{0}}\bigg)^{\sigma } \notag \\
&\times &(1+x)\frac{\rho _{-\tau }}{\rho }\bigg[1+(\sigma-1)\frac{\rho _{\tau }}{\rho }\bigg].
\label{MDIBpr}
\end{eqnarray}%

Similarly, the $B$-term in Eq. (\ref{EsymMDIEq}) for the symmetry
energy, i.e.,
\begin{eqnarray}
E_{\rm sym,B} = - \frac{B x}{\sigma + 1}\left(\frac{\rho}{\rho_0}\right)^\sigma
\label{EsymMDIBEq}
\end{eqnarray}
changes to
\begin{eqnarray}
E'_{\rm sym,B} = \frac{B}{\sigma + 1}\left(\frac{\rho}{\rho_0}\right)^\sigma\bigg[\frac{1-x}{4}\sigma(\sigma + 1)-\frac{1+x}{2}\bigg],
\label{EsymMDIBEqpri}
\end{eqnarray}
while the $B$-term in Eq.~(\ref{UsymMDIEq}) for the symmetry
potential, i.e.,
\begin{eqnarray}
&&U_{\rm sym,B}=-2x\frac{B}{\sigma+1}\Big(\frac{\rho}{\rho_0}\Big)^\sigma
\label{UsymMDIBEq}
\end{eqnarray}
changes to
\begin{eqnarray}
&&U'_{\rm sym,B}=\frac{B}{\sigma + 1}\left(\frac{\rho}{\rho_0}\right)^\sigma\bigg[\frac{1-x}{2}\sigma(\sigma + 1)-1-x\bigg].
\label{UsymMDIBEqpri}
\end{eqnarray}
The $x$-dependent parameters $A_{u}(x)$ and $A_{l}(x)$ then become,
respectively,
\begin{equation}
A_u'(x)=A_{u0} + \frac{2B}{\sigma+1}\bigg[\frac{1-x}{4}\sigma(\sigma + 1)-\frac{1+x}{2}\bigg],
\label{AuMDIBEq}
\end{equation}
and
\begin{equation}
A_l'(x)=A_{l0} - \frac{2B}{\sigma+1}\bigg[\frac{1-x}{4}\sigma(\sigma + 1)-\frac{1+x}{2}\bigg].
\label{AlMDIBEq}
\end{equation}
It should be noted that the expressions in Eqs.~(\ref{EsymMDIBEqpri}), (\ref{AuMDIBEq}),
and (\ref{AlMDIBEq}) are different from the corresponding ones in Ref.~\cite{XuC10a} where
the parabolic approximation Eq.~(\ref{EsymPA}) has been used to calculate the symmetry
energy while here the symmetry energy is obtained from the definition of Eq.~(\ref{Esym}).

\begin{figure}[htb]
\centering
\includegraphics[scale=0.88]{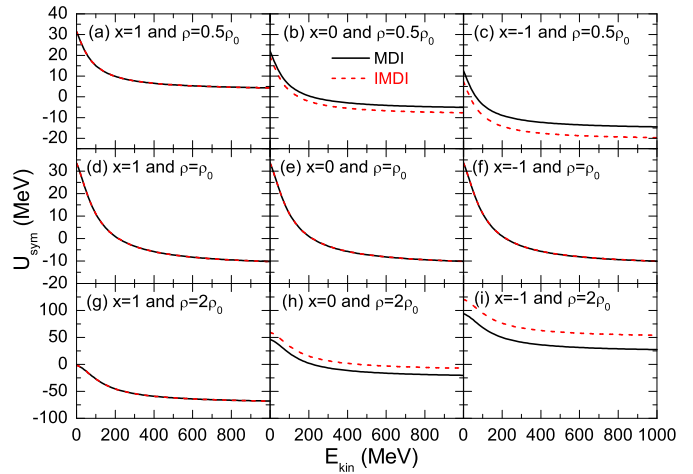}
\caption{The nuclear symmetry potential as a function of the nucleon
kinetic energy from both the MDI interaction (solid lines) and the
IMDI interaction (dashed lines) at density of
$\rho = \rho_0 /2$, $\rho_0$, and $2\rho_0$.}
\label{UsymEkinIMDI}
\end{figure}

\begin{figure}[htb]
\centering
\includegraphics[scale=0.9]{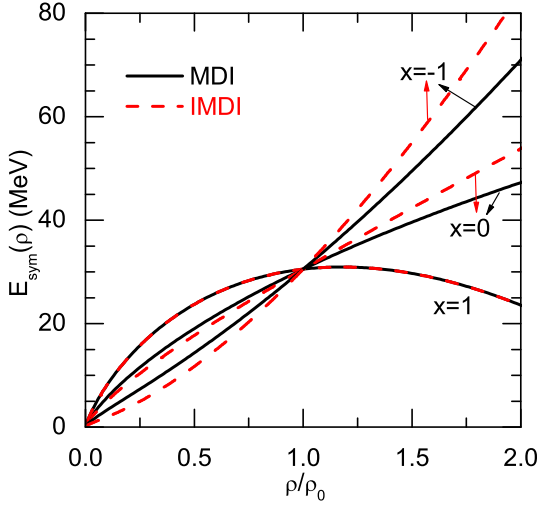}
\caption{Density dependence of the symmetry energy in the MDI
interaction (solid lines) and the IMDI interaction (dashed lines)
with $x=1$, $0$, and $-1$.} \label{EsymRhoIMDI}
\end{figure}

To illustrate the effects of the separate density dependence for
neutrons and protons, we show in Fig.~\ref{UsymEkinIMDI} the kinetic
energy dependence of the symmetry potential at density $\rho =
\rho_0 /2$, $\rho_0$, and $2\rho_0$, using the improved MDI (IMDI)
interaction. The results from the MDI interaction are also
included for comparison. Three typical values of the $x$ parameter,
i.e., $x=1$, $0$, and $-1$ are used. It is seen from panels (a),
(d), and (g) that the symmetry potentials from the MDI and IMDI
are exactly the same for $x=1$ due to the fact that
Eq.~(\ref{UsymMDIBEqpri}) is reduced to Eq.~(\ref{UsymMDIBEq}) for
$x=1$. Furthermore, one can see from panels (d), (e), and (f)
that the symmetry potentials from the MDI and IMDI
interactions are also exactly the same at $\rho = \rho_0$ for
different $x$ values due to construction (See, e.g., Eqs.
(\ref{AuMDIBEq}) and (\ref{AlMDIBEq})). For the cases with $x=0$ and
$-1$, one can see from panels (b), (c), (h), and (i) that the
symmetry potential from the IMDI interaction deviates significantly
from the one from the MDI interaction at $\rho \neq \rho_0$. Therefore,
one can expect that the symmetry energy from the MDI and
IMDI interactions for both $x=0$ and $-1$ will give significantly
different density behaviors for the symmetry energy, and this
indeed can be seen from Fig.~\ref{EsymRhoIMDI} where the density
dependence of the symmetry energies from the MDI and IMDI interactions
are compared. For $x=1$, as expected, one can see that the symmetry
energy is the same for both the MDI and IMDI interactions since
the $U_{\rm sym}(\rho,p)$ remains unchanged for $x=1$ (see
Fig.~\ref{UsymEkinIMDI}). For $x=0$ and $-1$, however, one can see
from Fig.~\ref{EsymRhoIMDI} that the symmetry energy from the IMDI
interaction becomes significantly stiffer compared to that from
the MDI interaction. These features are consistent with the
variations of the symmetry potential $U_{\rm sym}(\rho,p)$ obtained
from the MDI and IMDI interactions as shown in Fig.~\ref{UsymEkinIMDI}.

\section{Summary and outlook}
\label{summary}

We have reviewed the isospin- and momentum-dependent MDI interaction
that has been developed during recent years. The MDI interaction
has been systematically and successfully applied in the study of the
transport model simulations of heavy ion collisions induced by
neutron-rich nuclei, the thermal properties of asymmetric nuclear
matter including the liquid-gas phase transition, and the properties of
neutron stars. Through comparing with the available experimental data,
these studies based on the MDI interaction provide, via a phenomenological
way, important information and understanding on the in-medium nuclear
effective interaction, especially its isospin and momentum dependence.
And this further puts important constraints on the EOS of asymmetric
nuclear matter, especially the density dependence of the symmetry
energy.

We have emphasized the importance of the momentum dependence of the
nuclear mean-field potential in nuclear medium, including its
isovector symmetry potential. The IBUU04 transport model simulations
of heavy ion collisions with the MDI interaction indicate that the
momentum dependence of the nucleon isoscalar potential and isovector
symmetry potential plays an important role in the degree of isospin
diffusion and the energy dependence of t/$^3$He ratio in heavy ion
collisions. In fact, the IBUU04 transport model analysis on the
isospin diffusion data from NSCL-MSU has already put important
constraints on the density dependence of the symmetry energy. The
momentum dependence of the nucleon isocalar potential and isovector
symmetry potential also plays an important role in understanding the
thermal properties of asymmetric nuclear matter. This has been
demonstrated from the studies on the temperature dependence of the
symmetry energy and symmetry free energy, liquid-gas phase
transition of asymmetric nuclear matter, and especially the novel
differential isospin fractionation phenomenon in asymmetric nuclear
matter. Moreover, the isospin- and momentum-dependent MDI
interaction has been successfully used to explore the core-crust
transition of neutron stars, and important constraints on the
transition density and pressure at the inner edge of neutron stars
have been obtained through using the MDI interaction parameters
constrained by terrestrial laboratory data.

Furthermore, we have introduced two extensions of the MDI
interaction that have been made recently. One is the extended MDI
interaction for the baryon octet and the other is the improved MDI
interaction with separate density dependence for neutrons and
protons. The latter takes into account more accurately the isospin
dependence of the in-medium many-body interactions and thus may
provide a more physically reasonable density functional for the
isospin dependence of the in-medium nuclear effective interaction.
As for the extended MDI interaction for the baryon octet, we would
like to emphasize that it is useful not only in understanding
the properties of hybrid stars as discussed in the present paper,
but also in investigating strangeness production and its isospin
effects in heavy ion collisions induced by neutron-rich nuclei at
higher energies (a few GeV/nucleon), and thus is potentially
important for exploring the high density behaviors of the symmetry
energy and the in-medium effective interactions between
nucleon-hyperon and hyperon-hyperon.

Although the MDI interaction has been used for different studies
as described in this paper, it still can be improved in some aspects.
For example, the MDI interaction predicts an attractive isoscalar
single-particle potential at high momenta/energies in symmetric
nuclear matter at saturation density, which is significantly weaker
than the empirical optical potential (see, e.g., Fig.~\ref{SPPrho0H})
when the nucleon momentum is larger than about $550$ MeV/c (i.e.,
the nucleon kinetic energy of about $160$ MeV). This is due to the
fact that the parameters of the MDI interaction are obtained from
fitting the single-particle potential from the Gogny-D1 interaction
that fails to describe the high-momentum/energy behaviors of the
empirical optical model potential, although it can give a good
description of the properties of finite nuclei. Therefore, it will
be interesting to re-fit the values of the parameters in the MDI
interaction to obtain a reasonable high energy behaviors for the isoscalar
single-particle potential and study its effects in transport model
simulations of heavy ion collisions at higher incident energies
(larger than about $200$ MeV/nucleon). In addition, the MDI interaction
has been mainly used so far to explore the isospin-dependent properties
of nuclear matter, and it will be very interesting to see how the MDI
interaction describe the spin-dependent properties of asymmetric nuclear
matter, especially the density and asymmetry at which the ferromagnetic
transition can occur in an asymmetric nuclear matter.
These works  are in progress and will be reported elsewhere.

Finally, we would like to point out that the high baryon density
and/or large isospin asymmetry reached in heavy ion collisions
induced by neutron-rich nuclei provide a unique experimental
condition to probe the isospin- and momentum-dependent effective
interaction in high density asymmetric nuclear matter. Indeed, this
has already been demonstrated from studying the isospin effects in
heavy ion collisions induced by neutron-rich nuclei based on the
transport model simulations with the MDI interaction. In addition,
the high baryon density and large isospin asymmetry expected in
the interior of neutron stars provide another excellent astrophysical
condition to probe the isospin- and momentum-dependent effective
interaction in high density asymmetric nuclear matter, and this has
been demonstrated from studying the properties of hybrid stars with
the extended MDI interaction for the baryon octet.

\section*{Acknowledgments}
This work was supported by the National Natural Science Foundation
of China under Grant No. 11275125, 11135011, 11175085, 11235001, and
11035001, the Shanghai Rising-Star Program under Grant No.
11QH1401100, the ``Shu Guang" project supported by Shanghai Municipal
Education Commission and Shanghai Education Development Foundation,
the Program for Professor of Special Appointment (Eastern Scholar)
at Shanghai Institutions of Higher Learning, the Science and
Technology Commission of Shanghai Municipality (11DZ2260700), the
Project Funded by the Priority Academic Program Development of
Jiangsu Higher Education Institutions (PAPD), the ``100-talent plan"
of Shanghai Institute of Applied Physics under grant Y290061011 from
the Chinese Academy of Sciences, the US National Science Foundation
under Grant No. PHY-1068572 and PHY-1068022, the Welch Foundation
under Grant No. A-1358, the US National Aeronautics and Space
Administration under grant NNX11AC41G issued through the Science
Mission Directorate, and the CUSTIPEN (China-U.S. Theory Institute
for Physics with Exotic Nuclei) under DOE grant number
DE-FG02-13ER42025.


\begin{thebibliography}{99}

\bibitem{LiBA98} B.A. Li, C.M. Ko, and W. Bauer, topical review, Int. Jour.
Mod. Phys. E \textbf{7}, (1998) 147.

\bibitem{LiBA01b} Isospin Physics in Heavy-Ion Collisions at Intermediate
Energies, Eds. Bao-An Li and W. Udo Schr\"{o}der (Nova Science Publishers,
Inc, New York, 2001).

\bibitem{Lat00} J.M. Lattimer and M. Prakash, Phys. Rep. \textbf{333}, 121
(2000); Science \textbf{304}, (2004) 536.

\bibitem{Dan02a} P. Danielewicz, R. Lacey, and W.G. Lynch, Science \textbf{%
298}, (2002) 1592.

\bibitem{Bar05} V. Baran, M. Colonna, V. Greco, and M. Di Toro, Phys. Rep.
\textbf{410}, (2005) 335.

\bibitem{Ste05a} A.W. Steiner, M. Prakash, J.M. Lattimer, and P.J. Ellis,
Phys. Rep. \textbf{411}, (2005) 325.

\bibitem{CKLY07} L.W. Chen, C.M. Ko, B.A. Li, and G.C. Yong, Front. Phys.
China \textbf{2}, (2007) 327 [arXiv:0704.2340].

\bibitem{LCK08} B.A. Li, L.W. Chen, and C.M. Ko, Phys. Rep. \textbf{464},
(2008) 113.

\bibitem{Bru67} K.A. Brueckner, S.A. Coon and J. Dabrowski, Phys. Rev.
\textbf{168}, (1967) 1184.

\bibitem{Sie70} P.J. Siemens, Nucl. Phys. \textbf{A141}, (1970) 225.

\bibitem{Sjo74} O. Sj\"oberg, Nucl. Phys. \textbf{A222}, (1974) 161.

\bibitem{Bom91} I. Bombaci and U. Lombardo, Phys. Rev. C \textbf{44}, (1991) 1892.

\bibitem{Zuo02} W. Zuo, A. Lejeune, U. Lombardo, J. F. Mathiot, Eur. Phys.
J. A \textbf{14}, (2002) 469.

\bibitem{Har87} B. ter Haar and R. Malfliet, Phys. Rep. \textbf{149},
(1987) 207.

\bibitem{Mut00} H. M\"uther and A. Polls, Prog. Part. Nucl. Phys. \textbf{45},
(2000) 243.

\bibitem{Dic04} W.H. Dickhoff and C. Barbieri, Prog. Part. Nucl. Phys.
\textbf{52}, (2004) 377.

\bibitem{Fri81} B. Friedman and V.R. Pandharipande, Nucl. Phys. \textbf{A361},
(1981) 502.

\bibitem{Lag81} I.E. Lagaris and V.R. Pandharipande, Nucl. Phys. \textbf{A369},
(1981) 470.

\bibitem{Akm98} A. Akmal, V. R. Pandharipande, and D. G. Ravenhall, Phys.
Rev. C \textbf{58}, (1998) 1804.

\bibitem{Ser97} B.D. Serot and J.D. Walecka, Int. Jour. Mod. Phys. E \textbf{6}, (1997) 515.

\bibitem{Fur04} R.J. Furnstahl, Lect. Notes Phys. \textbf{641}, (2004) 1.

\bibitem{Vre04} D. Vretenar and W. Weise, Lect. Notes Phys. \textbf{641}, (2004) 65.

\bibitem{Ser86} B.D. Serot and J.D. Walecka, Adv. Nucl. Phys. \textbf{16},
(1986) 1.

\bibitem{Chi77} S.A. Chin, Ann. Phys. (N.Y.), \textbf{108}, (1977) 301.

\bibitem{Rei89} P.-G. Reinhard, Rep. Prog. Phys. \textbf{52}, (1989) 439.

\bibitem{Rin96} P. Ring, Prog. Part. Nucl. Phys. \textbf{37}, (1996) 193.

\bibitem{Vau72} D. Vautherin and D. M. Brink, Phys. Rev. C \textbf{5},
(1972) 626.

\bibitem{Bra85} M. Brack, C. Guet and H. -B. Hakansson, Phys. Rep.
\textbf{123}, (1985) 275.

\bibitem{Sto07} J.R. Stone and P.-G. Reinhard, Prog. Part. Nucl. Phys.
\textbf{58}, (2007) 587.

\bibitem{Tre86} J. Treiner \textit{et al.}, Ann. Phys. (N.Y.), \textbf{170},
(1986) 406.

\bibitem{Ban90} D. Bandyopadhyay, C. Samanta, S.K. Samaddar and J.N. De,
Nucl. Phys. \textbf{A511}, (1990) 1.

\bibitem{Bas03} D.N. Basu, Phys. Lett. \textbf{B566}, (2003) 90.

\bibitem{Bas05} D.N. Basu, P.R. Chowdhury, C. Samanta, Phys. Rev. C \textbf{72}, (2005) 051601(R).

\bibitem{Muk07} T. Mukhopadhyay and D. N. Basu, Nucl. Phys. \textbf{A789}, (2007) 201.

\bibitem{Bas08} D.N. Basu, P. Roy Chowdhury, and C. Samanta, Nucl. Phys. \textbf{A811}, (2008) 140.

\bibitem{Bas09} D.N. Basu, P. Roy Chowdhury, and C. Samanta, Phys. Rev. C \textbf{80}, (2009) 057304.

\bibitem{Cho09} P. Roy Chowdhury, D.N. Basu, and C. Samanta, Phys. Rev. C \textbf{80}, (2009) 011305(R).

\bibitem{Rou11} T.R. Routray, S.K. Tripathy, B.B. Dash, B. Behera, and D.N. Basu, Eur. Phys. J. A \textbf{47}, (2011) 92.

\bibitem{Rou12} T.R. Routray, X. Vinas, S.K. Tripathy, M. Bhuyan, S.K. Patra, and B. Behera, arXiv:1208.4236.

\bibitem{Gai13} T. Gaitanos and M. Kaskulov, Nucl. Phys. \textbf{A899}, (2013) 133.

\bibitem{Bas13} D.N. Basu, arXiv:1309.6793, 2013.

\bibitem{Die03} A.E.L. Dieperink, Y. Dewulf, D. Van Neck, M. Waroquier, and
V. Rodin, Phys. Rev. C \textbf{68}, (2003) 064307.

\bibitem{LiZH06} Z.H. Li, U. Lombardo, H.-J. Schulze, W. Zuo, L.W. Chen, and
H.R. Ma, Phys. Rev. C\textbf{74}, (2006) 047304.

\bibitem{Cai12} B.J. Cai and L.W. Chen, Phys. Rev. C \textbf{85}, (2012) 024302.

\bibitem{Zha01} F.S. Zhang and L.W. Chen, Chin. Phys. Lett. \textbf{18}, (2001) 142.

\bibitem{Ste06} A.W. Steiner, Phys. Rev. C \textbf{74}, (2006) 045808.

\bibitem{Xu09a} J. Xu, L.W. Chen, B.A. Li, and H.R. Ma, Phys. Rev. C \textbf{79}, (2009) 035802.

\bibitem{Xu09b} J. Xu, L.W. Chen, B.A. Li, and H.R. Ma, Astrophys. J. \textbf{697}, (2009) 1549.

\bibitem{ChenLW11b} L.W. Chen, Science China: Physics, Mechanics and Astronomy \textbf{54}, (2011) s124 [arXiv:1101.2384].

\bibitem{ZhangZ13} Z. Zhang and L.W. Chen, Phys. Lett. \textbf{B726}, (2013) 234.

\bibitem{XuC11} C. Xu, B. A. Li, L. W. Chen, and C. M. Ko, Nucl. Phys. A {\bf 865}, (2011) 1.

\bibitem{ChenR12} R. Chen, B.J. Cai, L.W. Chen, B.A. Li, X.H. Li, and C. Xu,
Phys. Rev. C \textbf{85}, (2012) 024305.

\bibitem{Lan62} A. M. Lane, Nucl. Phys. \textbf{35}, (1962) 676.

\bibitem{LiBA04b} B.A. Li, Phys. Rev. C 69, (2004) 034614.

\bibitem{XuC10} C. Xu, B. A. Li, and L. W. Chen, Phys. Rev. C \textbf{82}, (2010)
054607.

\bibitem{Cai12b} B.J. Cai and L.W. Chen, Phys. Lett. \textbf{B711}, (2012) 104.

\bibitem{LiXH13} X. H. Li, B. J. Cai, L. W. Chen, R. Chen, B. A. Li, and C.
Xu, Phys. Lett. \textbf{B 721}, (2013) 101.

\bibitem{XuC13} C. Xu, B.A. Li, and L.W. Chen, contribution in this volume [arXiv:1308.1502].

\bibitem{Ber88b} G.F. Bertsch and S. Das Gupta, Phys. Rep. \textbf{160}, (1988) 189.

\bibitem{Beh79} B. Behera and R.K. Satpathy, J. Phys. G \textbf{5}, (1979) 85.

\bibitem{Dec80} J. Decharge and D. Gogny, Phys. Rev. C \textbf{21} (1980) 1568.

\bibitem{Wir88} R. Wiringa, Phys. Rev. C \textbf{38}, (1988) 2967.

\bibitem{Cse92} L.P. Csernai, G. Fai, C. Gale, and E. Osnes, Phys. Rev. C \textbf{46}, (1992) 736.

\bibitem{Beh97} B. Behera, T.R. Routray, and R.K. Satpathy, J. Phys. G \textbf{23}, (1997) 445.

\bibitem{Beh98} B. Behera, T.R. Routray, and R.K. Satpathy, J. Phys. G \textbf{24}, (1998) 2073.

\bibitem{Gal87} C. Gale, G. Bertsch, and S. Das Gupta, Phys. Rev. C \textbf{35}, (1987) 1666.

\bibitem{Wel88} G.M. Welke, M. Prakash, T. T. S. Kuo, S. Das Gupta, and C. Gale,
Phys. Rev. C \textbf{38}, (1988) 2101.

\bibitem{Bom01} I. Bombaci, in Ref.~\cite{LiBA01b}, p.35.

\bibitem{Das03} C.B. Das, S. Das Gupta, C. Gale, and B.A. Li, Phys. Rev. C
\textbf{67}, (2003) 034611.

\bibitem{Che05a} L.W. Chen, C.M. Ko, and B.A. Li, Phys. Rev. Lett. \textbf{94}, (2005) 032701.

\bibitem{XuC10a} C. Xu and B.A. Li, Phys. Rev. C \textbf{81}, (2010) 044603.

\bibitem{Sto03} J.R. Stone, J.C. Miller, R. Koncewicz, P.D. Stevenson,
M.R. Strayer, Phys. Rev. C \textbf{68}, (2003) 034324.

\bibitem{XuC10b} C. Xu and B.A. Li, Phys. Rev. C \textbf{81}, (2010) 064612.

\bibitem{XuJ10b} J. Xu and C.M. Ko, Phys. Rev. C \textbf{82}, (2010) 044311.

\bibitem{Beh02} B. Behera, T.R. Routray, B. Sahoo, and R.K. Satpathy, Nucl. Phys. \textbf{A699}, (2002) 770.

\bibitem{Beh05} B. Behera, T.R. Routray, A. Pradhan, S.K. Patra, and P.K. Sahu, Nucl. Phys. \textbf{A753}, (2005) 367.

\bibitem{Beh07} B. Behera, T.R. Routray, A. Pradhan, S.K. Patra, and P.K. Sahu, Nucl. Phys. \textbf{A794}, (2007) 132.

\bibitem{Beh11} B. Behera, T.R. Routray, and S K Tripathy, J. Phys. G \textbf{38}, (2011) 115104.

\bibitem{LiBA04a} B.A. Li, C. B. Das, S. Das Gupta, and C. Gale, Phys. Rev.
C \textbf{69}, (2004) 011603(R); Nucl. Phys. \textbf{A735}, (2004) 563.

\bibitem{Che04} L.W. Chen, C.M. Ko, and B.A. Li, Phys. Rev. C \textbf{69}, (2004) 054606.

\bibitem{LiBA05a} B.A. Li, G.C. Yong and W. Zuo, Phys. Rev. C \textbf{71}, (2005) 014608.

\bibitem{LiBA05b} B.A. Li, G.C. Yong and W. Zuo, Phys. Rev. C \textbf{71}, (2005) 044604.

\bibitem{LiBA05c} B.A. Li and L.W. Chen, Phys. Rev. C \textbf{72}, (2005) 064611.

\bibitem{LiBA06b} B.A. Li, L.W. Chen, G.C. Yong, and W. Zuo, Phys. Lett. \textbf{B634}, (2006) 378.

\bibitem{Yon06a} G.C. Yong, B.A. Li, L.W. Chen, and W. Zuo, Phys. Rev. C \textbf{73}, (2006) 034603.

\bibitem{Yon06b} G.C. Yong, B.A. Li, L.W. Chen, Phys. Rev. C \textbf{74}, (2006) 064617.

\bibitem{Yon07} G.C. Yong, B.A. Li, and L.W. Chen, Phys. Lett. \textbf{B650}, (2007) 344.

\bibitem{Xu07} J. Xu, L.W. Chen, B.A. Li and H.R. Ma, Phys. Rev. C \textbf{75}, (2007) 014607.

\bibitem{Xu07b} J. Xu, L.W. Chen, B.A. Li and H.R. Ma, Phys. Lett. \textbf{B650}, (2007) 348.

\bibitem{Xu08} J. Xu, L.W. Chen, B.A. Li and H.R. Ma, Phys. Rev. C \textbf{77}, (2008) 014302.

\bibitem{Che07} L.W. Chen, C.M. Ko, and B.A. Li, Phys. Rev. C \textbf{76}, 054316 (2007).

\bibitem{Zuo06} W. Zuo, U. Lombardo, H.-J. Schulze, and Z.H. Li, Phys. Rev. C \textbf{74}, (2006) 014317.

\bibitem{Dal05} E.N.E. van Dalen, C. Fuchs, and A. Faessler, Phys. Rev. C \textbf{72}, (2005) 065803.

\bibitem{Che05c} L.W. Chen, C.M. Ko, and B.A. Li, Phys. Rev. C \textbf{72}, (2005) 064606.

\bibitem{LiZH06b} Z.H. Li, L.W. Chen, C.M. Ko, B.A. Li, and H.R. Ma, Phys.
Rev. C \textbf{74}, (2006) 044613.

\bibitem{Mur87} D.P. Murdock and C.J. Horowitz, Phys. Rev. C \textbf{35}, (1987) 1442.

\bibitem{McN83} J.A. McNeil, L. Ray, and S.J. Wallace, Phys. Rev. C \textbf{27}, (1983) 2123.

\bibitem{Jam89} M. Jaminon and C. Mahaux, Phys. Rev. C \textbf{40}, (1989) 354.

\bibitem{Neg98} J.W. Negele and H. Orland, Quantum Many-Particle System,
Perseus Books Publishing, L.L.C., 1998.

\bibitem{Fuc04} E.N.E. van Dalen, C. Fuchs and A. Faessler, Nucl. Phys. \textbf{A741}, (2004) 227;
Phys. Rev. Lett. \textbf{95}, (2005) 022302.

\bibitem{Sjo76} O. Sj\"oberg, Nucl. Phys. \textbf{A265}, (1976) 511.

\bibitem{Ma04} Z.Y. Ma, J. Rong, B.Q. Chen, Z.Y. Zhu and H.Q. Song, Phys.
Lett. \textbf{B604}, (2004) 170.

\bibitem{Zuo05} W. Zuo, L.G. Gao, B.A. Li, U. Lombardo and C.W. Shen, Phys.
Rev. C \textbf{72}, (2005) 014005.

\bibitem{Sam05} F. Sammarruca, W. Barredo and P. Krastev, Phys. Rev. C \textbf{71}, (2005) 064306.

\bibitem{LiXH12} X.H. Li and L.W. Chen, Nucl. Phys. A \textbf{874}, (2012) 62.

\bibitem{Neg81} J.W. Negele and K. Yazaki, Phys. Rev. Lett. \textbf{62}, (1981) 71.

\bibitem{Pan91} V.R. Pandharipande and S.C. Pieper, Phys. Rev. C \textbf{45}, (1991) 791.

\bibitem{LiGQ94} G.Q. Li and R. Machleidt, Phys. Rev. C \textbf{48}, (1994) 1702; {\it ibid}, C {\bf 49}, (1994) 566.

\bibitem{Sam05b} F. Sammrruca and P. Krastev, nucl-th/0506081, 2005.

\bibitem{Ber88} G.F. Bertsch, G.E. Brown, V. Koch, and B.A. Li, Nucl. phys. \textbf{A490}, (1988) 745.

\bibitem{Mao94} G.J. Mao, Z.X. Li, Y.Z. Zhuo, Y.L. Han, and Z.Q. Yu,
Phys. Rev. C \textbf{49}, (1994) 3137; G.G. Mao, Z.X. Li and Y.Z. Zhuo,
{\it ibid}, C \textbf{53}, (1996) 2933; C \textbf{55}, (1997) 792.

\bibitem{Cai05} T. Caitanos, C. Fuchs and H.H. Wolter, Phys. Lett \textbf{B609}, (2005) 241.

\bibitem{Xiao09}  Z.G. Xiao, B.A. Li, L.W. Chen, G.C. Yong, and M. Zhang, Phys. Rev. Lett. \textbf{102}, (2009) 062502.

\bibitem{Zha09} M. Zhang et al.,  Phys. Rev. C \textbf{80}, (2009) 034616.

\bibitem{Zha10} M. Zhang et al.,  Phys. Rev. C \textbf{82}, (2010) 044602.

\bibitem{Yon08} G.C. Yong, B.A. Li and L.W. Chen, Phys. Lett. \textbf{B661}, (2008) 82.

\bibitem{Xiao13}  Z.G. Xiao, G.C. Yong, L.W. Chen, B.A. Li, G.Q. Xiao, and N. Xu, contribution in this volume [arXiv:1312.5790].

\bibitem{Tsa04} M.B. Tsang et al., Phys. Rev. Lett. \textbf{92}, (2004) 062701.

\bibitem{Ram00} F. Rami et al., Phys. Rev. Lett. \textbf{84}, (2000) 1120.

\bibitem{Ber84} G.F. Bertsch, H. Kruse and S. Das Gupta, Phys. Rev. C \textbf{29}, (1984) 673.

\bibitem{Shi03} L. Shi and P. Danielewicz, Phys. Rev. C \textbf{68}, (2003) 064604.

\bibitem{LiBA97a} B.A. Li, C.M. Ko, Z.Z. Ren, Phys. Rev. Lett. \textbf{78}, (1997) 1644.

\bibitem{Che03b} L.W. Chen, C.M. Ko, B.A. Li, Phys. Rev. C \textbf{68}, (2003) 017601; Nucl. Phys. \textbf{A729}, (2003) 809.

\bibitem{Hod03} P.E. Hodgson and E. B\u{e}t\'{a}k, Phys. Rep. \textbf{374}, (2003) 1; and the references therein.

\bibitem{Cse86} L.P. Csernai and J.I. Kapusta, Phys. Rep. \textbf{131}, (1986) 223; and the references therein.

\bibitem{Gyu83} M. Gyulassy, K. Frankel, and E.A. Relmer, Nucl. Phys. \textbf{A402}, (1983) 596.

\bibitem{Aich87} J. Aichelin, A. Rosenhauer, G. Peilert, H. St\"{o}cker, and W. Greiner, Phys. Rev. Lett. \textbf{58}, (1987) 1926.

\bibitem{Koch90} V. Koch \textit{et al.}, Phys. Lett. \textbf{B241}, (1990) 174.

\bibitem{Indra00} P. Pawlowski \textit{et al.}, Eur. Phys. Jour. A \textbf{9}, (2000) 371.

\bibitem{Mat95} R. Mattiello \textit{et al.}, Phys. Rev. Lett. \textbf{74}, (1995) 2180;
R. Mattiello \textit{et al.}, Phys. Rev. C \textbf{55}, (1997) 1443.

\bibitem{Nag96} J. L. Nagle \textit{et al.}, Phys. Rev. C \textbf{53}, (1996) 367.

\bibitem{Pol99} A. Polleri et al., Nucl. Phys. \textbf{A661}, (1999) 452c.

\bibitem{Hagel00} K. Hagel \textit{et al.}, Phys. Rev. C \textbf{62}, (2000) 034607.

\bibitem{Cibor00} J. Cibor \textit{et al.}, Phys. Lett. \textbf{B473}, (2000) 29.

\bibitem{Sobotka01} L.G. Sobotka, R.J. Charity, and J.F. Dempsey, in Ref.~\cite{LiBA01b}, p.331.

\bibitem{Vesel01} M. Veselsky \textit{et al.}, Phys. Lett. \textbf{B497}, (2001) 1.

\bibitem{Chomaz99} Ph. Chomaz and F. Gulminelli, Phys. Lett. \textbf{B447}, (1999) 221.

\bibitem{Che01} L.W. Chen, F.S. Zhang, Z.H. Lu, W.F. Li, Z.Y. Zhu,
and H.R Ma, J. Phys. G \textbf{27}, (2001) 1799.

\bibitem{Zuo03} W. Zuo et al., Phys. Rev. C \textbf{69},
(2003) 064001; \textit{ibid}. C \textbf{73}, (2006) 035208.

\bibitem{LiBA06} B.A. Li and L.W. Chen, Phys. Rev. C \textbf{74}, (2006) 034610.

\bibitem{Mou07} Ch. C. Moustakidis, Phys. Rev. C \textbf{76}, (2007) 025805.

\bibitem{De12} J.N. De and S.K Samaddar, Phys. Rev. C \textbf{85}, (2012) 024310.

\bibitem{Don94} P. Donati, P.M. Pizzochero, P.F. Bortignon, and R.A.
Broglia, Phys. Rev. Lett. \textbf{72}, (1994) 2835.

\bibitem{Dea95} D.J. Dean, S.E. Koonin, K. Langanke, and P.B. Radha,
Phys. Lett. \textbf{B356}, (1995) 429; D.J. Dean, K. Langanke, and J.M.
Sampaio, Phys. Rev. C \textbf{66}, (2002) 045802.

\bibitem{Gal90} C. Gale, G. M. Welke, M. Prakash, S. J. Lee, and S.
Das Gupta, Phys. Rev. C \textbf{41}, (1990) 1545.

\bibitem{Lee01} S.J. Lee and A.Z. Mekjian, Phys. Rev. C \textbf{63},
(2001) 044605.

\bibitem{Mek05} A.Z. Mekjian, S.J. Lee, and L. Zamick, Phys. Rev. C \textbf{72}, (2005) 044305.

\bibitem{betty01} M.B. Tsang {\it et al.}, Phys. Rev. Lett. \textbf{86}, (2001) 5023.

\bibitem{shetty} D. V. Shetty et al., Phys. Rev. C \textbf{70},
(2004) 011601(R); D. V. Shetty et al., \textit{ibid}. C \textbf{71},
(2005) 024602; D.V. Shetty et al., arXiv:nucl-ex/0603016.

\bibitem{sjy} G.A. Souliotis et al., Phys. Rev. C \textbf{73},
(2006) 024606; J. Iglio et al., \textit{ibid}. C \textbf{74}, (2006) 024605;
G. A. Souliotis et al., arXiv:nucl-ex/0603006.

\bibitem{shetty06} D.V. Shetty et al., arXiv:nucl-ex/0606032.

\bibitem{indra} A. Le F\'evre et al. for the ALADIN and INDRA
collaborations, Phys. Rev. Lett. \textbf{94}, (2005) 162701.

\bibitem{wolfgang} W. Trautmann et al. for the ALADIN and INDRA
collaborations, Proceedings of the IWM2005, Catania, Italy, Nov 2005 [arXiv:nucl-ex/0603027].

\bibitem{kowalski} S. Kowalski et al., Phys. Rev. C \textbf{75}, (2007) 014601.

\bibitem{tsang01} M.B. Tsang et al., Phys. Rev. C \textbf{64}, (2001) 054615.

\bibitem{botvina} A.S. Botvina, O.V. Lozhkin and W. Trautmann, Phys.
Rev. C \textbf{65}, (2002) 044610.

\bibitem{ono} A. Ono, P. Danielewicz, W.A. Friedman, W.G. Lynch and
M.B. Tsang, Phys. Rev. C \textbf{68}, (2003) 051601(R); \textit{ibid}. C
\textbf{70}, (2004) 041604; arXiv:nucl-ex/0507018.

\bibitem{dorso} C.O. Dorso, C.R. Escudero, M. Ison, and J.A. L\'{o}%
pez, Phys. Rev. C \textbf{73}, (2006) 044601.

\bibitem{ma} Y.G. Ma et al., Phys. Rev. C \textbf{69},
(2004) 064610; Y.G. Ma et al., \textit{ibid}. C \textbf{72}, (2005) 064603; W.D.
Tian et al., Chin. Phys. Lett. \textbf{22}, (2005) 306.

\bibitem{horowitz} C.J. Horowitz and A. Schwenk, Nucl. Phys. \textbf{A776}, (2006) 55.

\bibitem{Nat10} J.B. Natowitz et al., Phys. Rev. Lett. \textbf{104}, (2010) 202501.

\bibitem{Typ13} S. Typel, H.H. Wolter, G. Ropke, and D. Blaschke, contribution in this Volume [arXiv:1309.6934v1].

\bibitem{lamb78} D. Q. Lamb, J. M. Lattimer, C. J. Pethick, and D.
G. Ravenhall, Phys. Rev. Lett. \textbf{41}, (1978) 1623.

\bibitem{fin82} J.E. Finn et al., Phys. Rev. Lett. \textbf{49}, (1982) 1321.

\bibitem{ber83} G.F. Bertsch and P.J. Siemens, Phys. Lett. \textbf{B126}, (1983) 9.

\bibitem{jaqaman83} H. Jaqaman, A. Z. Mekjian, and L. Zamick, Phys.
Rev. C \textbf{27}, (1983) 2782; \textit{ibid}. C \textbf{29}, (1984) 2067.

\bibitem{chomaz} Ph. Chomaz, M. Colonna, and J. Randrup, Phys. Rep.
\textbf{389}, (2004) 263.

\bibitem{das} C.B. Das, S. Das Gupta, W.G. Lynch, A.Z. Mekjian, and
M.B. Tsang, Phys. Rep. \textbf{406}, (2005) 1.

\bibitem{wci} \textit{Dynamics and Thermodynamics with Nucleonic
Degrees of Freedom}, Eds. Ph. Chomaz, F. Gulminelli, W. Trautmann and S.J.
Yennello, Springer, (2006).

\bibitem{Mul95} H. M\"{u}ller and B.D. Serot, Phys. Rev. C
\textbf{52}, (1995) 2072.

\bibitem{Su00} W.L. Qian, R.K. Su, and P. Wang, Phys. Lett. \textbf{B491}, (2000) 90.

\bibitem{Xu00} H.S. Xu \textit{et al.}, Phys. Rev. Lett. \textbf{85}, (2000) 716.

\bibitem{LiBA07} B.A. Li, L.W. Chen, H.R. Ma, J. Xu, and G.C. Yong, Phys. Rev. C
\textbf{76}, (2007) 051601(R).

\bibitem{Rav83} D.G. Ravenhall, C.J. Pethick, and J.R. Wilson, Phys. Rev.
Lett. \textbf{50}, (1983) 2066.

\bibitem{Oya93} K. Oyamatsu, Nucl. Phys. \textbf{A561}, (1993) 431.

\bibitem{Hor04} C.J. Horowitz et al., Phys. Rev. C \textbf{69}, (2004) 045804;
C.J. Horowitz et al., Phys. Rev. C \textbf{70}, (2004) 065806.

\bibitem{Ste08} A.W. Steiner, Phys. Rev. C \textbf{77}, (2008) 035805.

\bibitem{Pet95b} C. J. Pethick, D. G. Ravenhall and C. P. Lorenz, Nucl. Phys. \textbf{A584}, (1995) 675.

\bibitem{Dou00} F. Douchin and P. Haensel, Phys. Lett. \textbf{B485},
(2000) 107.

\bibitem{Dou01} F. Douchin and P. Haensel, A\&A \textbf{380}, (2001) 151.

\bibitem{Hor03} J. Carriere, C.J. Horowitz, and J. Piekarewicz, Astrophys. J. \textbf{593}, (2003) 463.

\bibitem{Kub07} S. Kubis, Phys. Rev. C \textbf{76}, (2007) 035801; Phys.
Rev. C \textbf{70}, (2004) 065804.

\bibitem{Lat07} J.M. Lattimer and M. Prakash, Phys. Rep. \textbf{442},
(2007) 109.

\bibitem{Wor08} A. Worley, P.G. Krastev, and B.A. Li, Astrophys. J. \textbf{685}, (2008) 390.

\bibitem{BPS71} G. Baym, C. Pethick, and P. Sutherland, Astrophys. J. \textbf{170}, (1971) 299.

\bibitem{BBP71} G. Baym, H.A. Bethe, and C.J. Pethick, Nucl. Phys. \textbf{A175}, (1971) 225.

\bibitem{Pet95a} C.J. Pethick and D.G. Ravenhall, Ann. Rev. Nucl. Part.
Sci. \textbf{45}, (1995) 429.

\bibitem{Oya07} K. Oyamatsu and K. Iida, Phys. Rev. C \textbf{75},
(2007) 015801.

\bibitem{Duc07} C. Ducoin, Ph. Chomaz, and F. Gulminelli, Nucl. Phys. \textbf{A789}, (2007) 403.

\bibitem{Pro06} C. Provid\^{e}ncia, L. Brito, S.S. Avancini, D.P. Menezes,
and P. Chomaz, Phys. Rev. C \textbf{73}, (2006) 025805.

\bibitem{Duc08a} C. Ducoin, J. Margueron, and P. Chomaz, Nucl. Phys. \textbf{A809}, (2008) 30.

\bibitem{Duc08b} C. Ducoin, C. Provid\^{e}ncia, A.M. Santos, L. Brito, and
P. Chomaz, Phys. Rev. C \textbf{78}, (2008) 055801.

\bibitem{Pai10} H. Pais, A. Santos, L. Brito, and C. Provid\^{e}ncia, Phys.
Rev. C \textbf{82}, (2010) 025801.

\bibitem{Hor01} C.J. Horowitz and J. Piekarewicz, Phys. Rev. Lett. \textbf{86%
}, (2001) 5647; Phys. Rev. C \textbf{64}, (2001) 062802(R); Phys.
Rev. C \textbf{66}, (2002) 055803.

\bibitem{Lin99} B. Link, R.I. Epstein, and J.M. Lattimer, Phys. Rev. Lett.
\textbf{83}, (1999) 3362.

\bibitem{Arp72} J. Arponen, Nucl. Phys. \textbf{A191}, (1972) 257.

\bibitem{ChenLW12} L.W. Chen, Proceedings of the 14th National Conference on Nuclear
Structure in China (NSC2012), Eds. J Meng, C.W. Shen, E.G. Zhao, S.G. Zhou
(World Scientific, Singapore, 2012), pp. 43-54 [arXiv:1212.0284].

\bibitem{Lat91} J.M. Lattimer, C.J. Pethick, M. Prakash, and P. Haensel, Phys. Rev. Lett. \textbf{66}, (1991) 2701.

\bibitem{XuJ10a} J. Xu, L.W. Chen, C.M. Ko, and B.A. Li, Phys. Rev. C \textbf{81}, (2010) 055803.

\bibitem{Has06} O. Hashimoto and H. Tamura, Prog. Part. Nucl. Phys. \textbf{57},
(2006) 564.

\bibitem{Dover89} C.B. Dover, D.J. Millener, and A. Gal, Phys. Rep.
\textbf{184}, (1989) 1.

\bibitem{Bart99} S. Bart, et al., Phys. Rev. Lett. \textbf{83}, (1999) 5238.

\bibitem{Dover83} C.B. Dover and A. Gal, Ann. Phys. \textbf{146}, (1983) 309.

\bibitem{Maessen89} P.M.M Maessen, Th. A. Rijken, and J.J. de Swart,
Phys. Rev. C \textbf{40}, (1989) 2226.

\bibitem{Reuber94} A. Reuber, K. Holinda, and J. Speth, Nucl. Phys. \textbf{A570}, (1994) 543.

\bibitem{Dabrowski99} J. Dabrowski, Phys. Rev. C \textbf{60}, (1999) 025205.

\bibitem{Millener88} D.J. Millener, C.B. Dover, and A. Gal, Phys.
Rev. C \textbf{38}, (1988) 2700.

\bibitem{Chrien89} R.E. Chrien and C.B. Dover, Ann. Rev. Nucl. Part.
Sci. \textbf{39}, (1989) 227.

\bibitem{Aok95} S. Aoki, et al., Phys. Lett. \textbf{B355}, (1995) 45.

\bibitem{Fukuda98} T. Fukuda, et al., Phys. Rev. C \textbf{58}, (1998) 1306.

\bibitem{Khaustov00} P. Khaustov, et al., Phys. Rev. C \textbf{61}, (2000) 054603.

\bibitem{Batty94} C.J. Batty, E. Friedman, and A. Gal, Phys. Lett. \textbf{B335}, (1994) 273.

\bibitem{Mares95} J. Mare$\check{s}$, E. Friedman, A. Gal, and B.K.
Jennings, Nucl. Phys. \textbf{A594}, (1995) 311.

\bibitem{Noumi02} H. Noumi, et al., Phys. Rev. Lett. \textbf{89}, (2002) 072301.

\bibitem{Har05} T. Harada and Y. Hirabayashi, Nucl. Phys. \textbf{A759}, (2005) 143.

\bibitem{Friedman07} E. Friedman and A. Gal, Phys. Rep.
\textbf{452}, (2007) 89.

\bibitem{Schaffer96} J. Schaffner and I.N. Mishustin, Phys. Rev. C
\textbf{53}, (1996) 1416.

\bibitem{Schaffer94} J. Schaffner, C.B. Dover, A. Gal, M. Hanauske, C. Greiner,
D.J. Millener, and H. St$\ddot{o}$cker, Ann. Phys. \textbf{235},
(1994) 35.

\bibitem{Dap09} H. Dapo, B. J. Schaefer, and J. Wambach, Phys. Rec. C \textbf{81}, (2010) 035803.

\bibitem{Pol06} H. Polinder, J. Haidenbauer, and U. G. Mei{\ss}ner,
Nucl. Phys. \textbf{A779}, (2006) 244.

\bibitem{Hama90} S. Hama, et al., Phys. Rev. C \textbf{41}, (1990) 2737.

\bibitem{Coo93} E. D. Cooper, S. Hama, B. C. Clark, and R. L. Mercer, Phys. Rev. C \textbf{47}, (1993) 297.

\bibitem{Baldo98} M. Baldo, G.F Burgio, and H.J. Schulze, Phys. Rev. C
\textbf{58}, (1998) 3688.

\bibitem{QFLi05} Q. Li, Z. Li, E. Zhao, and R. K. Gupta,
Phys. Rev. C \textbf{71}, (2005) 054907.

\bibitem{Glen92} N.K. Glendenning, Phys. Rev. D
\textbf{46}, (1992) 1274.

\bibitem{Glen01} N.K. Glendenning, Phys. Rep.
\textbf{342}, (2001) 393.

\bibitem{Chodos74} A. Chodos, R.L. Jaffe, K. Johnson, C.B Thorn, and
V.F. Weisskopf, Phys. Rev. D \textbf{9}, (1974) 3471.

\bibitem{Heinz86} U. Heinz, P.R. Subramanian, H. Stocker, and W.
Greiner, J. Phys. G \textbf{12}, (1986) 1237.

\bibitem{Bal00} M. Baldo, G.F. Burgio, and H.-J. Schulze,
Phys. Rev. C \textbf{61}, (2000) 055801.

\bibitem{Sam10} F. Sammarruca, Int. J. Mod. Phys. E \textbf{19}, (2010) 1259.

\bibitem{koh} S. K\"ohler, Nucl. Phys. \textbf{A258}, (1976) 301.

\bibitem{Dut1} E. Chabanat et al., Nucl. Phys. \textbf{627}, (1997) 710.

\bibitem{Dut2} A. K. Dutta, J.-P. Arcoragi, J. M. Pearson, R. Behrman and E
Tondeur, Nucl. Phys. \textbf{A458}, (1986) 77.

\bibitem{Dut3} F. Tondeur, A. K. Dutta, J. M. Pearson and R. Behrman, Nucl. Phys. \textbf{A470}, (1987) 93.

\bibitem{Dut4} J. M. Pearson, Y. Aboussir, A. K. Dutta, R. C. Nayak,
M. Farine and E Tondeur, Nucl. Phys. \textbf{A528}, (1991) 1.

\bibitem{Dut5} Y. Aboussir, J. M. Pearson, A. K. Dutta and F. Tondeur, Nucl. Phys. \textbf{A549}, (1992)
155.

\bibitem{Dut6}  M. Farine, J. M. Pearson and E. Tondeur, Nucl. Phys. \textbf{A615}, (1997) 135.

\bibitem{bru64} K. A. Brueckner and J. Dabrowski, Phys. Rev. \textbf{134}, (1964) B722.

\bibitem{Dab73} J. Dabrowski and P. Haensel, Phys. Rev. C \textbf{7}, (1973) 916; Can. J. Phys. \textbf{52}, (1974) 1768.

\bibitem{Spr} D. W. L. Sprung and P. K. Banerjee, Nucl. Phys. \textbf{A168}, (1971) 273.

\bibitem{Neg} J. W. Negele, Phys. Rev. C \textbf{l}, (1970) 1260.

\end{thebibliography}
\end{document}